\documentclass[fleqn,usenatbib]{mnras}

\usepackage{newtxtext,newtxmath}

\usepackage[T1]{fontenc}
\usepackage{ae,aecompl}

\usepackage{aas_macros}
\usepackage{booktabs} 
\usepackage{siunitx}
\usepackage{subcaption}
\usepackage{tikz}

\usepackage{graphicx}	
\usepackage{amsmath}	
\usepackage{amssymb}	

\newcommand{\code}[1]{\texttt{#1}}
\newcommand{\lcdm}{$\Lambda$CDM}

\newcommand{\Msun}{\ensuremath{\mathrm{M}_\odot}}

\newcommand{\mean}[1]{\ensuremath{\left\langle #1 \right\rangle}}
\newcommand{\abs}[1]{\left\lvert#1\right\rvert}

\newcommand{\vect}[1] {\ensuremath{\bmath{#1}}}
\newcommand{\tens}[1] {\ensuremath{\mathbfss{#1}}} %

\newcommand{\hMpc}[1]{\ensuremath{{#1}\,h^{-1} \mathrm{Mpc}}}
\newcommand{\Mpch}[1]{\ensuremath{{#1}\,h \mathrm{Mpc}^{-1}}}
\newcommand{\hGpc}[1]{\ensuremath{{#1}\,h^{-1} \mathrm{Gpc}}}
\newcommand{\hMsun}[1]{\ensuremath{\num{#1}\,h^{-1} M_\odot}}
\newcommand{\Om}{\ensuremath{\Omega_\text{m}}}
\newcommand{\Ol}{\ensuremath{\Omega_\Lambda}}

\title[Haloes in MG and DE]{Dark matter haloes in modified gravity and dark energy:
  interaction rate,  small-, and large-scale alignment}

\author[B.~L'Huillier et al]
{Benjamin~L'Huillier,$^{1,2}$\thanks{E-mail:benjamin@kasi.re.kr}
  Hans~A.~Winther,$^{3,6}$
  David~F.~Mota,$^4$
  Changbom~Park,$^2$
  \newauthor
  and Juhan~Kim$^{5}$\\
  $^1$ Korea  Astronomy  and  Space Science Institute,  Yuseong-gu, 776
  Daedeok daero, 34055 Daejeon, Korea\\
  $^2$ School of Physics, Korea Institute for Advanced
  Study, 85 Hoegi-ro, Dongdaemun-gu, Seoul 130-722, Korea\\
  $^3$ Astrophysics,  University of  Oxford, DWB, Keble  Road, Oxford,
  OX1 3RH, United Kingdom\\
  $^4$ Institute for Theoretical Astrophysics, University of Oslo, Norway\\
  $^5$ Center for Advanced Computation, Korea Institute for Advanced
  Study, 85 Hoegi-ro, Dongdaemun-gu, Seoul 130-722, Korea\\
  $^6$ Institute of Cosmology \& Gravitation, University of Portsmouth, Portsmouth, Hampshire, PO 3FX, UK
}

\date{Accepted 2017 March 20. Received 2017 March 16; in original form 2016 October 24.}

\pubyear{2017}

\begin{document}
\label{firstpage}
\pagerange{\pageref{firstpage}--\pageref{lastpage}}
\maketitle

\begin{abstract}
  We study the properties of dark matter  haloes in a wide range of modified
  gravity models, namely, $f(R)$, DGP, and interacting dark energy models.
  We study the effects of modified gravity and dark energy on the internal properties of haloes, such as the spin and the structural parameters.
  We find that $f(R)$ gravity enhance the median value of the Bullock spin parameter, but  could not detect such effects for DGP and coupled dark energy.
  $f(R)$ also yields a lower median sphericity and oblateness, while coupled dark energy has the opposite effect. 
  However, these effects are very small.
	We then study the interaction rate of haloes in different gravity, and find that only strongly coupled dark energy models enhance the interaction rate. 
We  then quantify the enhancement of  the alignment of the  spins of interacting halo pairs by modified gravity.
  Finally, we study  the alignment of the major axes
  of haloes with the large-scale structures.
  The alignment of the spins  of interacting pairs of haloes in
  DGP  and coupled  dark energy  models show  no discrepancy  with GR,
  while $f(R)$ shows a weaker alignment.
  Strongly coupled dark energy shows  a stronger alignment of the halo
  shape with the large-scale structures. 
\end{abstract}




\section{Introduction}

The discovery of the acceleration of  the expansion of the Universe by
\citet{1998AJ....116.1009R}  and \citet{1998Natur.391...51P},  as well
as       studies      from       the      large-scale       structures
\citep[e.g.,  ][]{1999RSPTA.357..105C} led  to  the  emergence of  the
\lcdm\ model,  where the  Universe is  dominated by  dark energy
(DE), responsible  for the acceleration  of the expansion, and  a cold
dark matter (CDM) component that drives structure formation.

However, the nature of dark matter and  dark energy is one of the main
puzzles in modern  physics. While there are  many plausible candidates
for dark  matter, dark energy  poses more theoretical  challenges. The
simplest choice is dark energy being  due to vacuum energy in the form
of a cosmological  constant in Einstein's field  equations. This model
is in perfect  agreement with current observations, but  is plagued by
the fine-tuning problem. 

The next  to simplest option is  that dark energy is  dynamical, as in
quintessence models. The dark energy fields can also have interactions
with the  dark matter  sector giving rise  to interacting  dark energy
models \citep[coupled  quintessence,][]{2000PhRvD..62d3511A}. In these
scenarios there is no interaction  between dark energy and baryons, as such
the  constraints  coming  from   local  gravity  experiments  are  not
applicable. Nevertheless,  there are  several cosmological  bounds for
these    models   specially    coming    from   structure    formation
(\citealt{tomi,Mota:2007sz,Mota:2007zn,Pettorino:2013,lei},           see
\citealt{2006IJMPD..15.1753C} for a review on dark energy).

Another possibility is that general  relativity (GR) does not describe
gravity         properly         on        cosmological         scales
\citep{2012PhR...513....1C}.  To  get  around  the  tight  constraints
coming  from high  precision experiments  on  Earth and  in the  Solar
System    \citep{Will2006LRR.....9....3W,2014LRR....17....4W}   viable
modified gravity models 
must have some form of  screening mechanism \citep{Khoury2010} to hide
the modifications in the high  density regimes (relative to the cosmic
mean) where these  experiments have been performed. 

In the recent years, both analytical and numerical studies of modified
gravity   and   screening    mechanisms   have   become   increasingly
performed
\citep{koivisto,  Gannouji:2010fc, Li:2011b,Li:2011c,    Llinares:2013, 
Llinares:2013qbh,  isis, sandstad, 2015MNRAS.454.4208W}. 
Not just because of it being a possible dark energy
candidate,  but also  due  to the  fact  that we  are  finally in  the
position where  we can make precision,  percent level, tests of  GR on
cosmological scales  just as we have  done in the Solar  System in the
last century. By  studying alternatives to GR we can  find new ways of
testing gravity on  scales we have not tested it  before, for instance
via    the   study    of   the    internal   properties    of   haloes
\citep{2015MNRAS.452.3179S},         the          lensing         mass
\citep{2011PhRvL.107g1303Z},          haloes         in          voids
\citep{2012MNRAS.421.3481L},  or  the  3-points  correlation  function
\citep{2016A&A...592A..38S}. 

Studying halo formation in modified gravity (MG) is thus important for
two reasons. If gravity is not correctly described by GR, but by a modified theory, 
one needs to understand how modified gravity affects galaxy formation. 
Moreover,  galaxies and  galaxy clusters themselves  can be
used as a test of gravity.  

Future  weak  lensing  surveys, such  as  Euclid  \citep{lrr-2013-6,
  2011arXiv1110.3193L}, will provide strong 
constrains on GR. However, the intrinsic alignment of galaxies will be
a  source of  systematics.  It  is thus  important  to understand  the
alignment of galaxies in MG. Many  groups have studied halo and galaxy
alignment in  \lcdm\  \citep[e.g.][]{2007MNRAS.381...41H,
  2013ApJ...766L..15L, 
  2012MNRAS.427.3320C}  and  \citet{2013ApJ...763...28L}  studied  the
spin of dark matter haloes in modified gravity.

\citet{2017MNRAS.tmp..129L}  studied  the  effects   of  interactions  on  the
small-scale  alignment  of  haloes  and their  dependencies  with  the
environment,  which was  possible thanks  to the  large volume  of the
Horizon  Run 4  \citep{2015JKAS...48..213K}. The  interaction rate  of
halos in the \lcdm\ model was studied in \citet{2015MNRAS.451..527L}. 

In  this  paper,  we  use state-of-the  art  $N$-body  simulations  of
different  modified gravity  and dark  energy models,  namely, $f(R)$,
DGP, and coupled dark energy, and  study the internal properties of dark matter
haloes, such as their structural and spin parameters, and external properties,  
such as their interaction  rate, small-scale alignment
with their interacting neighbour, and  large-scale alignment with the
cosmic web.

The models  are described in  \S~\ref{sec:models}, \S~\ref{sec:method}
presents the simulations we used and the method.
\S~\ref{sec:intern}  deals  with  the internal  properties  of  haloes
(spins   and   structural  parameters),   \S~\ref{sec:inter}   studies
interacting pairs,  and \S~\ref{sec:lss}  is devoted to  the alignment
with the large-scale structures.

\section{Gravity models}
\label{sec:models}

The main ingredient in a successful modified gravity model is a screening mechanism that hides the modifications of gravity on Earth and in the Solar-System allowing it to pass the stringent constraints coming from local gravity experiments. The study of modified gravity models, which may or may not be particularly interesting in their own right, can be thought of as a way of studying how the underlying screening mechanism work. Several different screening mechanisms are known in the literature \citep{joyce} and in this paper we will consider two models, $f(R)$ gravity and the normal branch DGP model, that have two different screening mechanisms in play: the chameleon and Vainhstein mechanism. For the chameleon mechanism \citep{Khoury2010} screening depends on the local value of the gravitational potential while for the Vainshtein mechanism screening is a function of the local matter density. Different screening mechanisms operating on nonlinear scales may give rise to unique features. It is therefore highly  desirable  to explore observational consequences  that help expose these differences using physical observables in the  non-linear regime of structure formation. Below we will give a brief overview of the two models we consider in this paper. Both of these models have a background evolution that is either identical or very close to \lcdm\ so the differences in structure formation comes solely from the addition of a fifth-force that alters the growth of structures.

\subsection{$f(R)$ gravity}
In $f(R)$ gravity models  \citep{2010LRR....13....3D} the Ricci scalar
$R$ is augmented by a general function $f(R)$ given the action
\begin{align}
S = \int \sqrt{-g}{\rm d}^4x\frac{M_{\rm Pl}^2}{2}\left[R + f(R)\right],
\end{align}
where $g$ is the determinant of the metric $g_{\mu\nu}$ and $M_{\rm Pl} \equiv 1/\sqrt{8\pi G}$.

The particular  $f(R)$ model  studied in this  paper is  the so-called
Hu-Sawicky model \citep{2007PhRvD..76f4004H} which is defined by
\begin{align}
f(R) = -m^2\frac{c_1(-R/m^2)^n}{1+c_2(-R/m^2)^n},
\end{align}
where    $n,c_1,c_2$    are     dimensionless    numbers    satisfying
$\frac{c_1}{c_2}   =  \frac{\Omega_\Lambda}{\Omega_m}$   and  $m^2   =
\Omega_m H_0^2$. In the high curvature regime $|R| \gg m$ we can write
\begin{align}
f(R) \simeq  -\frac{f_{R0}}{n}
R_0\left(\frac{R_0}{R}\right)^n,
\end{align}
where  $R_0$  ($f_{R0}$)   is  the  present  value  of   $R$  ($f_R  =
\frac{df(R)}{dR}$) respectively. For all the simulations in this paper
$n=1$  so the  models are  defined  by a  single dimensionless  number
$f_{R0}$. 
For the range of $|f_{R0}|$ values that we consider in this paper the background evolution is almost indistinguishable from \lcdm.

$f(R)$  gravity can  via a  conformal transformation  be written  as a
scalar-tensor theory  where $f_R$ plays  the role of the  scalar field
\citep{2008PhRvD..78j4021B}. The equation determining the evolution of
$f_R$ is given by
\begin{align}
\nabla^2 f_R = -\frac{a^2}{3}\left[\delta R +
  \frac{\delta\rho_m}{M_{\rm Pl}^2}\right] =
-\frac{a^2}{3}\left[\sqrt{\frac{f_{R0}}{f_R}}R_0 - R(a)  +
  \frac{\delta\rho_m}{M_{\rm Pl}^2}\right],
\end{align}
where $\delta  R = R - R(a)$ and $R(a)$  is the background  value for
$R$. In an $N$-body simulation of $f(R)$ gravity this equation is solved
at  every time-step  to determine  the fifth-force  $\frac{1}{2}\nabla
f_R$.

The parameter $f_{R0}$ controls the  range of the  scalar interaction
and in the cosmological background today we have
\begin{align}
\lambda_{0} = 7.46\cdot \sqrt{\frac{f_{R0}}{10^{-5}}} ~\hMpc{}.
\end{align}
Roughly speaking on large length  scales $r \gtrsim \lambda_0$ gravity
behaves  as  General Relativity  while  on  small scales  $r  \lesssim
\lambda_0$  gravity  is  modified.  In  addition to  this  we  have  a
screening  effect, the chameleon mechanism, in  high density  regions. For  objects that  have a
large Newtonian potential $\Phi_N$ the  fifth-force is suppressed by a
factor   $\left|\frac{3f_{R}}{2\Phi_N}\right|$.  Thus   the  parameter
$f_{R0}$ also  acts as  a critical potential;  objects at  the present
time  with $|\Phi_N|  \gg |f_{R0}|$  do not  feel any  modification of
gravity  while  objects  with  $|\Phi_N|  \lesssim  |f_{R0}|$  feel  a
modified Newton's constant $G_{\rm eff} = \frac{4}{3}G$.

\subsection{DGP model}

The DGP  \citep{2000PhLB..485..208D} model  is a  so-called braneworld
model where our  Universe is confined to a 4D  brane which is embedded
in a 5D spacetime. The action is given by 
\begin{align}
S  =  \int\sqrt{-g_{(4)}}{\rm d}^4x\frac{R_{(4)}}{2}M_{\rm Pl~(4)}^2    +
\int\sqrt{-g_{(5)}}{\rm d}^5x\frac{R_{(5)}}{2}M_{\rm Pl~(5)}^2,
\end{align}
where $g_{(4)}$  ($g_{(5)}$) denotes the  metric on the brane  (in the
bulk) and $R_{(4)}$ ($R_{(5)}$) denotes  the Ricci scalar on the brane
(in the  bulk). Since  $M_{\rm Pl (4)}  = 1/\sqrt{8\pi  G}$ we
only have one  free parameter in the model which  is usually expressed
as the so-called cross-over scale $r_c = 
\frac{1}{2}\left(\frac{M_{\rm  Pl~(4)}}{M_{\rm Pl~(5)}}\right)^2$.

The  modifications of  the gravity  in the  model is  determined by  a
scalar-field $\phi$,  the brane-bending mode, which  described how the
brane   we   live   on   curves   in   the   fifth-dimension.   In   a
Friedmann-Lema\^itre-Robertson-Walker   background  the   gravitational
potential, which determines how particles move in 
a $N$-body   simulation,  is   given  by   $\Phi  =   \Phi_N  +
\frac{\phi}{2}$   where    $\Phi_N$   is   the    standard   Newtonian
potential. The dynamics of $\phi$ in the quasi-static approximation is
determined by 
\begin{align}
\nabla^2\phi   +   \frac{r_c^2}{3\beta  a^2}\left[(\nabla^2\phi)^2   -
  (\nabla_i\nabla_j\phi)^2\right] =  \frac{a^2\delta\rho}{\beta M_{\rm
    Pl}^2},
\end{align}
where
\begin{align}
\beta(a) = 1 + 2H(a)r_c\left(1 + \frac{\dot{H}(a)}{3H^2(a)}\right).
\end{align}
The model  we are  working with  here is the  normal branch  DGP model
apposed  to  the  original   DGP  model  which  had  self-accelerating
solutions.  The latter  one is  effectively  ruled out \citep{lrr-2010-5}. In  the
normal branch DGP model the acceleration  of the Universe is driven by
a cosmological constant just as in \lcdm. This model is a
useful  toy-model to  study  the particular  screening mechanism,  the
so-called Vainshtein mechanism, used by  DGP to hide the modifications
of gravity in  local experiments. The modifications of  gravity in the
vicinity of  a massive object of mass $M$ and radius $r_c$ are determined by  a scale known  as  the  Vainshtein  radius $r_V  \propto  r_c^{2/3}  M^{1/3}$.
Test-particles  far   outside  the  Vainshtein  radius   will  feel  a
gravitational    force   that    is   enhanced,    $G_{\rm   eff}    =
G_N\left(1+\frac{1}{3\beta(a)}\right)$, while test-particles far inside the  Vainshtein radius will just  feel the
standard Newtonian gravitational force $G_{\rm eff} \simeq G_N$. 
This basically means that we have screening in a region if the average matter density is higher than some critical value $\rho_{\rm crit} \simeq \frac{9\beta^2}{4(r_cH_0)^2}\rho_{c}$ where $\rho_{c} = 3H^2/8\pi G$ is the critical matter density in the Universe. For $\rho_m \gg \rho_{\rm crit}$ the fifth-force is suppressed by a factor $\approx \sqrt{\rho_{\rm crit}/\rho_m}$.

\subsection{Interacting dark energy}
In   interacting   dark  energy   models   \citep{1995A&A...301..321W,
  2000PhRvD..62d3511A}, the acceleration of
the  Universe is  driven by  a (quintessence)  scalar field  which has
interactions with dark matter, leading  to energy exchange between the
two  fluids as  the Universe  expands. Baryons and  radiation are  not
coupled to the scalar field as such a coupling is strongly constrained
by  solar system  tests  of gravity  requiring $\beta_{\rm  baryons}^2
\lesssim 10^{-5}$. The dynamical equations at the background 
level are given by
\begin{align}
\ddot{\phi}    +    3H\dot{\phi}     +    \frac{dV(\phi)}{d\phi}    &=
\sqrt{\frac{2}{3}}\beta(\phi)\frac{\rho_{\rm DM}}{M_{\rm Pl}},\\
\dot{\rho}_{\rm       DM}       +       3H\rho_{\rm       DM}       &=
-\sqrt{\frac{2}{3}}\beta(\phi)\dot{\phi}\frac{\rho_{\rm    DM}}{M_{\rm
    Pl}},\\
\dot{\rho}_{\rm b} + 3H\rho_{\rm b} &= 0.
\end{align}
Each model  on this form is  specified by a potential  $V(\phi)$ and a
coupling function $\beta(\phi)$.  The mass of the  scalar field $\phi$
is $\mathcal{O}(H_0)$ at  the present time which means  that the field
does  not cluster  significantly. The  gravitational force  on the  dark
matter particles is therefore  equivalent to a time-dependent Newton's
constant
\begin{align}
G_{\rm eff}^{\rm DM} = G\left(1 + \frac{4}{3}\beta^2(a)\right),
\end{align}
where  $\beta(a) =  \beta(\phi(a))$  and $\phi(a)$  is the  background
solution for $\phi$.  The absence of a coupling to  baryons means that
there is no need for a screening mechanism to be consistent with local
tests  of  gravity.  Observations  of the  CMB  places  the  strongest
constraints  on  the  model  which constraints  $\beta  \lesssim  0.2$
\citep{2016JCAP...01..045C}. 

For  the models  considered  here  \citep{2012MNRAS.422.1028B} we  have
$\beta(\phi)  =  \beta_0   e^{\frac{\beta_1  \phi}{M_{\rm  Pl}}}$  and
$V(\phi) = Ae^{-\frac{\alpha\phi}{M_{\rm Pl}}}$ where $A=0.00218M_{\rm
  Pl}^4,\alpha=0.08,\beta_1 = 0$ and $\beta_0 = 0.05$ (the EXP1 model)
and $\beta_0 = 0.15$ (the EXP3  model). The scalar field is normalized
such that $\phi(z=0) = 0$. 

\section{Methods}
\label{sec:method}

In this section we describe the  simulations and the generation of the
halo catalogues we have used for our analysis. 

\subsection{The simulations}

To study  the effects of  modified gravity  on galaxy haloes,  we took
advantage  of a  wide range  of available  simulations with  different
models.

In brief, we  used four different sets of simulations,  each set using
its own initial conditions and cosmology.
For each set of simulations, we use the \lcdm\ run as a reference
for the corresponding modified gravity runs.

\begin{table}
  \caption{\label{tab:sims}%
    Summary of the simulations used in this work}
  \begin{center}
    \begin{tabular}{lllll}
      \toprule
      Set & Name & Normal- & $L$ & $N$ \\
      & & ization & \hMpc{} & \\
      \midrule
      f512 & \lcdm\ & CMB & 256 & $512^3$\\
      &F4 & CMB & 256 & $512^3$\\
      &F5 & CMB & 256 & $512^3$\\
      &F6 & CMB & 256 & $512^3$\\
      \midrule
      f1024 & \lcdm & $\sigma_8=0.8 $ & 1024 & $ 1024^3$\\
      &F5 &$\sigma_8=0.8$ & 1024 & $1024^3 $\\
      \midrule
      DGP & \lcdm\ & CMB & 250 & $512^3$\\
      & DGP1.2& CMB & 250 & $512^3$\\
      \midrule
      CoDECS & \lcdm & CMB & 1000 & $1024^3$\\
      & EXP1 & CMB & 1000 & $1024^3$\\
      & EXP3 & CMB & 1000 & $1024^3$\\
      \bottomrule
    \end{tabular}
  \end{center}
\end{table}

The  first set  of simulations,  f512, uses  $512^3$ particles  in a
$L=\hMpc{256}$.
The reference \lcdm\ cosmology for this set of simulations is
$(\Om, \Ol ,h,\sigma_8) = (0.267,0.733,0.719, 0.80)$.
The  $f(R)$ simulations  have  $f_{R0} =  10^{-4}$ (F4),  $10^{-5}$
(F5) and $10^{-6}$ (F6).
The  initial density  fluctuations have  been normalized  by the  CMB,
which yields different $\sigma_8$ at $z=0$.

The second  set of  simulations, f1024, uses  $1024^3$ particles  in a
$L=\hMpc{1024}$ box. The reference cosmology is
$(\Om, \Ol ,h,\sigma_8) = (0.267,0.733,0.719, 0.80)$.
The associated MG simulations is $f(R)$ models with
$f_{R_0} =10^{-5}$, but has been normalized to the same
$\sigma_8 = 0.80$ at $z=0$.

The above two sets of simulations  were run using the \code{ISIS} code
\citep{2014A&A...562A..78L}, which is based on \code{RAMSES} 
\citep{2002A&A...385..337T}.
The code  used was recently  compared to other modified  gravity codes
\citep{2015MNRAS.454.4208W} and for the two models studied here, $f(R)$ and DGP, excellent (sub-percent) agreement was found in the enhancement of the matter matter power-spectrum relative to \lcdm\ (both computed within each code) for scales $k\lesssim \Mpch{5-10}$. Similar percent level agreement was found for the enhancement of the velocity divergence power spectrum, halo abundances and halo profiles for all redshifts studied $z \lesssim 2$.

The third set of simulation aims to study DGP gravity.
The reference GR  cosmology has $(\Omega_\text{m},\Omega_\Lambda,h) =
(0.271, 0.729,0.703)$,
with $512^3$ particles in a $L=\hMpc{250}$  box, and was also run with
\code{ISIS}. The DGP run was run with the crossover scale $r_c = 1.2
/ H_0$. 

The  fourth  set of  simulations  comes  from the  publicly  available
L-CoDECS  project \citep{2010MNRAS.403.1684B,  2012MNRAS.422.1028B}, a
set of $N$-body simulations of coupled dark energy, evolving
$2\times   1024^3$  particles   (dark  matter  and  baryons  without
hydrodynamics) in a  $L=\hMpc{1000}$ box  with the modified  version of \code{Gadget-2} \citep{2005MNRAS.364.1105S} by \citet{2010MNRAS.403.1684B}.
The reference cosmology is
$(\Om,  \Ol, h)  = (0.2711,  0.7289, 0.703)$. We  used two
simulations:  EXP1   and  EXP3,  which  have   an  exponential  scalar
self-interaction   potential   with   $\beta_0  =   0.05$   and   0.15
respectively. 
Note that, to fairly compare the CoDECS  simulations, which include baryons, with the $N$-body ones,
we exclude  the baryon particles from our the CoDECS simulations, 
and correct the mass of the CDM particles accordingly by $\Omega_\mathrm{m} / \Omega_\text{CDM}$.

We note  that, since each  set of  simulations (f512, f1024,  DGP, and
CoDECS) has different initial  conditions and cosmologies, the results
should  not be  compared  between the  sets, but  within  a given  set
between each model.  

\subsection{Halo detection and catalogues}

The haloes and subhaloes were respectively detected as in
\citet{2015MNRAS.451..527L}     using     the    Ordinary     Parallel
Friend-of-Friend (\code{OPFOF}) and physically self-bound (\code{PSB}) algorithms.
\citep{2006ApJ...639..600K}.
\code{OPFOF} is a  memory-efficient parallel implementation of  \code{FoF}, and \code{PSB}
is  a subhalo  finder that  finds density  peaks in  a similar  way to
\code{SUBFIND} \citep{2001MNRAS.328..726S},  and additionally  truncates the
subhaloes to the tidal radius.
For more details, we refer the reader to \citet{2006ApJ...639..600K}.

For  consistency  between different  gravity  models,  we defined  the
virial radius $R_\text{vir}$ as
 \begin{align}
 	\label{eq:virial}
   \frac {M(<R_\text{vir})}{4/3\pi R_\text{vir}^3} & = 200 
   \rho_\text{c}(z), \\
 \intertext{where}
 \rho_\text{c}(z) &= \frac{3H^2(z)}{8\pi G}
\end{align}
is the critical density of the Universe.\footnote{%
  This    definition   is    different    from   the    one   used    in
  \citet{2015MNRAS.451..527L}, and  does not take into  account the
  dependence   of    $\Delta_\text{c}$   with   redshift,    since   the
  \citet{1998ApJ...495...80B}  formula   is  only   valid  for   a  flat
  \lcdm\ cosmology.
}

\code{PSB} also calculates the potential energy  and the tidal radius of each
subhalo candidate before removing unbound particles.
In this step, we also assume a Newtonian gravity in all cases.
We expect this  to slightly underestimate the bound masses  in MG, by
removing particles that are actually bound to the halo.
This has been shown to slightly underestimate the mass function 
\citep{2010PhRvD..81j4047L}.

Targets  consist  of  \code{PSB}  sub-haloes with  more  than  50  particles,
yielding a minimal mass  of respectively \num{4.72e11}, \num{3.78e12},
\num{4.47e11}, and \hMsun{3.57e12} in the f512, f1024, DGP, and CoDECS
simulations. 
A  target is  defined to  be  interacting with  a neighbour  if it  is
located within  twice the virial radius  of its neighbour, and  if the
neighbour is at least 0.4 the mass of the target.
We note that  we slightly changed the definition  of interactions with
respect to \citet{2015MNRAS.451..527L, 2017MNRAS.tmp..129L} in order to increase 
our statistics. 

In order  to define the  large-scale density, we calculated  the density
field $\rho_{20}$ smoothed over the 20 nearest neighbours of each target halo
\citep{2015MNRAS.451..527L, 2017MNRAS.tmp..129L}, and defined as
\begin{align}
  \rho_{20} &= \sum_{i=1}^{20} M_i W(r_i,h), \text{where}\\
  W(r_i,h) &= \frac{1}{\pi h^3}
  \begin{cases} 
    1-\frac{3}{2}         \left(\frac{r}{h}\right)^2+        \frac{3}{4}
    \left(\frac{r}{h} \right)^3, & 0\leqslant \frac{r}{h} \leqslant 1\\ 
    \frac{1}{4}\left( 2-\left(\frac{r}{h}\right)\right)^3 , & 1\leqslant
    \frac{r}{h} \leqslant 2 \\ 
    0, & \frac{r}{h} \geqslant 2
  \end{cases}
\end{align}
is the cubic  spline kernel used in  smoothed particle
hydrodynamics   \citep[e.g.,][]{1985A&A...149..135M},  $r_i$   is  the
distance to the $i$th neighbour, and $h$ is 
the smoothing length, defined as the distance to the 21st neighbour.
The large-scale overdensity is thus defined as
\begin{equation}
  1+\delta_{20} = \frac{\rho_{20}}{\bar\rho_\text{h}},
\end{equation}
where $\bar\rho_\text{h}$ is the mean halo density. 

\section{Spin and shape of subhaloes in modified gravity}
\label{sec:intern}

In this section, we present our  results from a systematic analysis of
the shape and  spin parameters of \code{PSB} subhaloes in  simulations of the
modified gravity and coupled dark energy theories. 

\subsection{Shapes}
\label{sec:struct}

\begin{figure*}
\includegraphics[width=\textwidth]{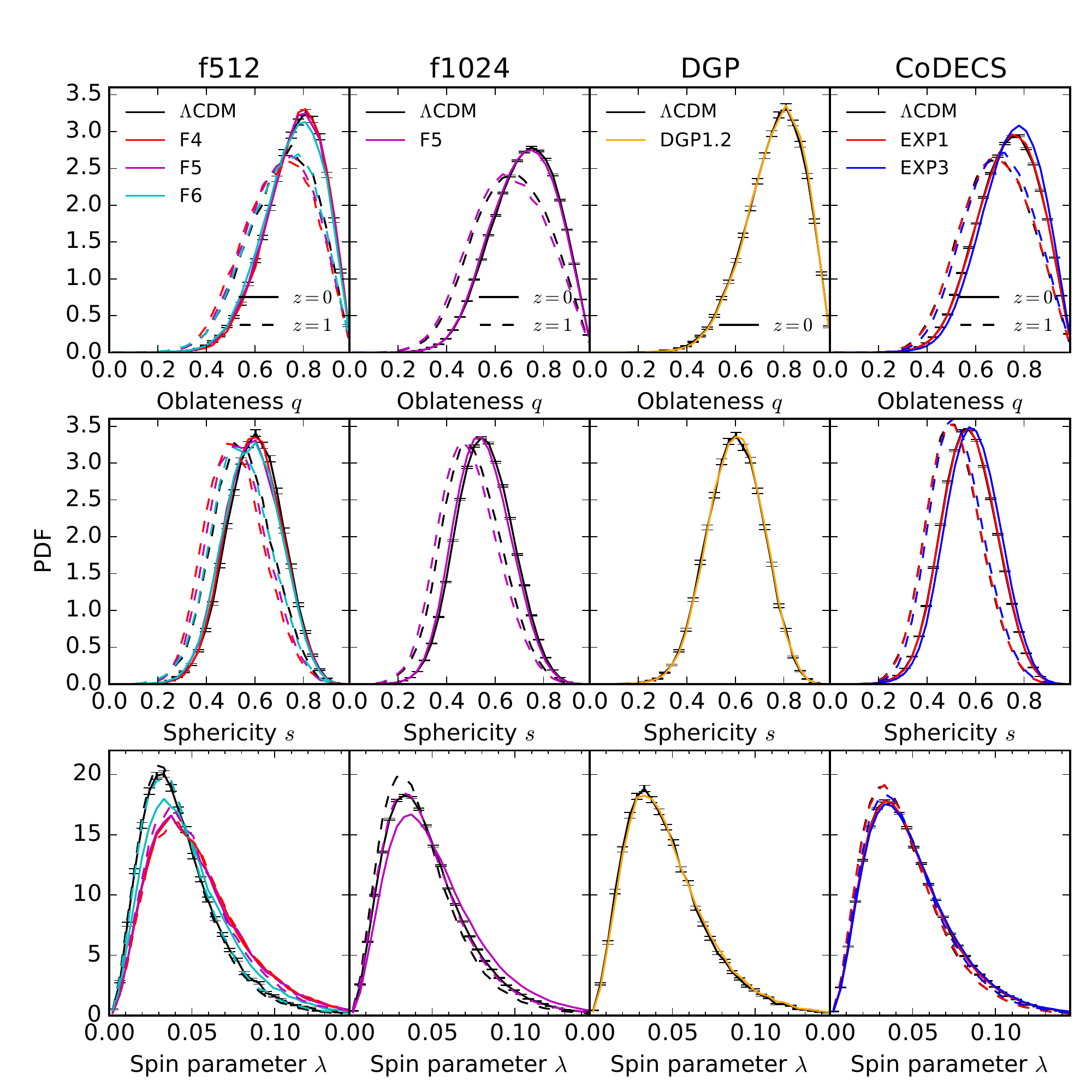}
\caption{\label{fig:pql}%
Distribution  of the  oblateness (top),  sphericity (middle)  and spin
(bottom) parameters in the 4 sets of simulations.
The  solid lines  show the  results at  $z=0$ and  the dashed  ones at
$z=1$. The \lcdm\ results are in black.
$f(R)$  yields  a  lower  sphericity,  oblateness,  and  Peebles  spin
parameter.
There is no sign for departure from GR in DGP.
Coupled  dark  energy  with  large  coupling  (EXP3)  yields  larger
oblateness and sphericity than GR, but  show no deviation for the spin
parameters. 
}
\end{figure*}

The internal  distribution of matter  can be described by  the inertia
tensor
\begin{align}
  \tens{I}_{ij} = \sum_\alpha x_{\alpha,i}x_{\alpha,j},
\end{align}
where $x_{\alpha,i}$ is the $i$th coordinate of particle $\alpha$.

The sphericity $s$ and  oblateness $q$ are defined as $q  = \frac b a$
and  $s =  \frac  c  a$ where  $a^2>b^2>c^2$  are  the eigenvalues  of
\tens{I}. We note that, in order  to limit resolution effects, we only
considered subhaloes resolved with more than 100 particles. 

The first and second rows  of Fig.~\ref{fig:pql} respectively show the
distribution of the oblateness $q$ and sphericity $s$ of \code{PSB} subhaloes
in the f512 (first column), f1024 (second column), DGP (third column),
and CoDECS (fourth column) sets of simulations.
The solid lines  show the distributions at $z=0$, and  the dashed lines
at $z=1$. The error-bars show the Poisson error in each bin and the \lcdm\ reference simulation in each set is shown in black.

In the first and second rows, $f(R)$  simulations  seem to  have  a  slightly lower  sphericity  and
oblateness, especially  at $z=1$. In  the   case  of   $f(R)$  gravity,   the  difference   between  the
distributions is larger at $z=1$ than $z=0$. In the DGP simulation we find no difference on the distribution of
the shape parameters compared to \lcdm. In interacting dark energy models,  the (weakly coupled) EXP1 model is
indistinguishable  from GR,  while  the (more  strongly coupled)  EXP3
model has a slightly larger median sphericity and oblateness. 
However, the shift is very small. For instance, in the case of CoDECS, the median and 68\% percentile of the oblateness $q$  are: $0.7561^{+  0.0589}_{-  0.0629}$ and $0.7406^{+  0.0615}_{-  0.0652}$ for EXP3 and \lcdm\ at $z=0$.

\subsection{Spin parameter}

\label{sec:spin_psb}

To describe the rotation of  haloes, we calculated the spin parameters
of each \code{PSB} subhalo, defined as  \citep{2001ApJ...555..240B} 
\begin{align}
  \label{eq:lambdab}
  \lambda_\text{B} &= \frac{\abs{\vect{J}}}{\sqrt 2 M R V},\\
  \intertext{where} 
  V^2 & = \frac{G M} R,
\end{align}
and $\vect J$ is the sum of  the angular momenta of each particle in
the  halo.

In the third row of  Fig.~\ref{fig:pql}, we show the spin distribution
of the spin parameters.
We only consider haloes resolved with more than 100 particles.
In $f(R)$ gravity, the fifth force tends to speed up halo rotation, in agreement with \citet{2013ApJ...763...28L}.
On the other hand, DGP and coupled dark energy seem to have no effect on the spin distribution. 

However, the effect of modified gravity on the distribution of the structural parameters is overall very small.
For instance, the median, 16$^\text{th}$, and 84$^\text{th}$ percentiles associated with the spin parameter at $z=0$ in the
 f1024 set is $0.0425^{+  0.0122}_{-  0.0101}$ and 
 $0.0467^{+  0.0136}_{-  0.0112}$ for the \lcdm\ and F5 models respectively, making the effect of modified gravity on the spin distribution difficult to detect observationally.

\begin{figure}
	\centering
	\includegraphics[width=\columnwidth]{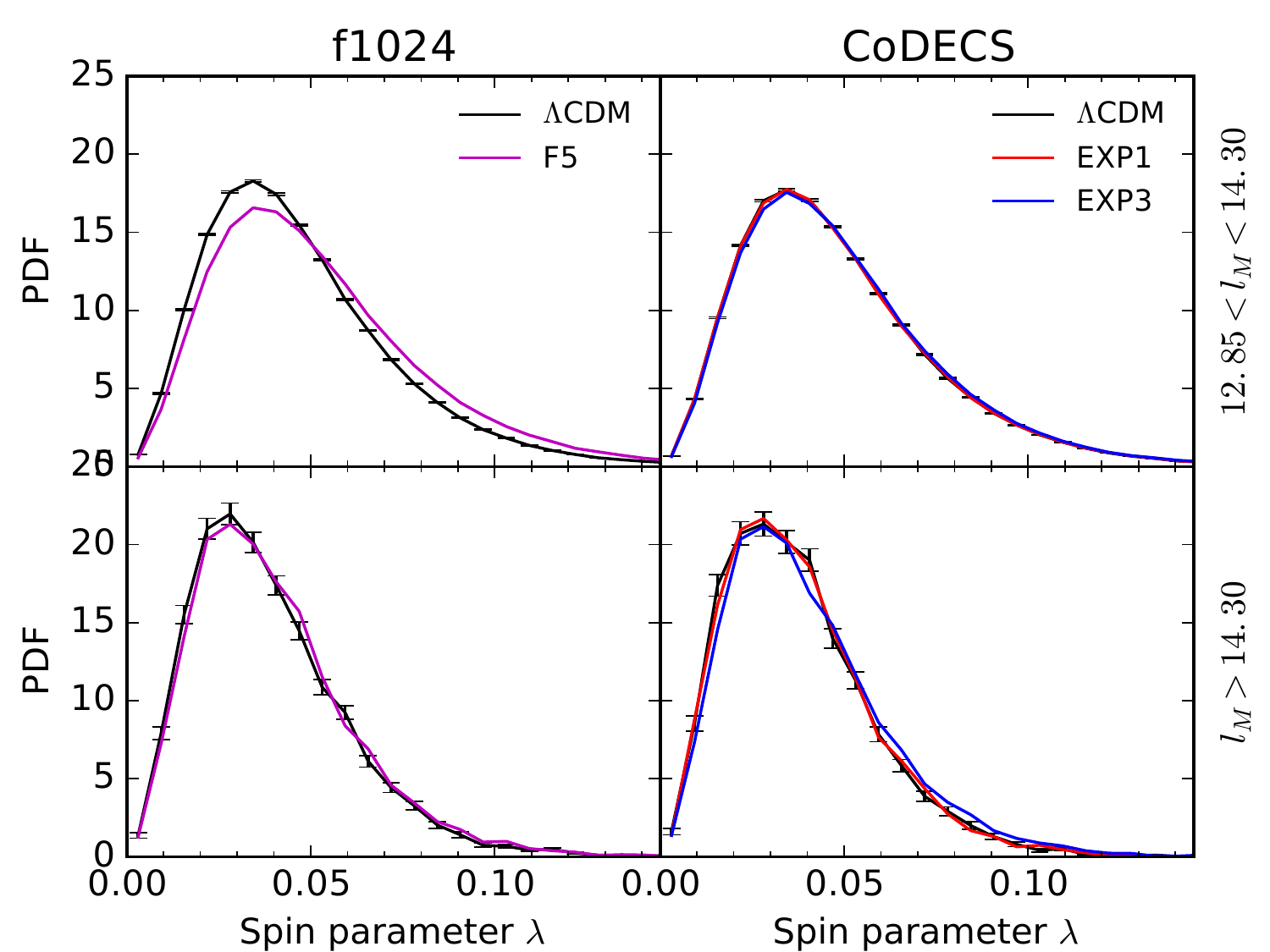}
	\caption{\label{fig:lmass}%
    	Spin distribution at $z=0$ for $l_M<14.30$ (top) and $l_M>14.30$ (bottom), where $l_M = \log_{10} hM/\Msun$.
    }
\end{figure}

Fig.~\ref{fig:lmass} shows the mass dependence of the spin distribution at $z=0$ in two mass bins, namely, 
$l_M< 14.30$ (top), and $l_M > 14.30$ (bottom), where 
$l_M = \log_{10} hM/\Msun$.
In the CoDECS simulations, as expected, the coupled dark energy models show no difference with \lcdm\ at any mass bins. 
However, in the lower mass bin, the spin distribution of F5 is shifted towards larger spins with respect to that of \lcdm, while no difference can be seen in the higher mass bin. 
Again, this can be understood as an effect of screening of the fifth force at high masses.

\begin{figure}
  \centering
  \begin{subfigure}{\columnwidth}
    \includegraphics[width=\columnwidth]{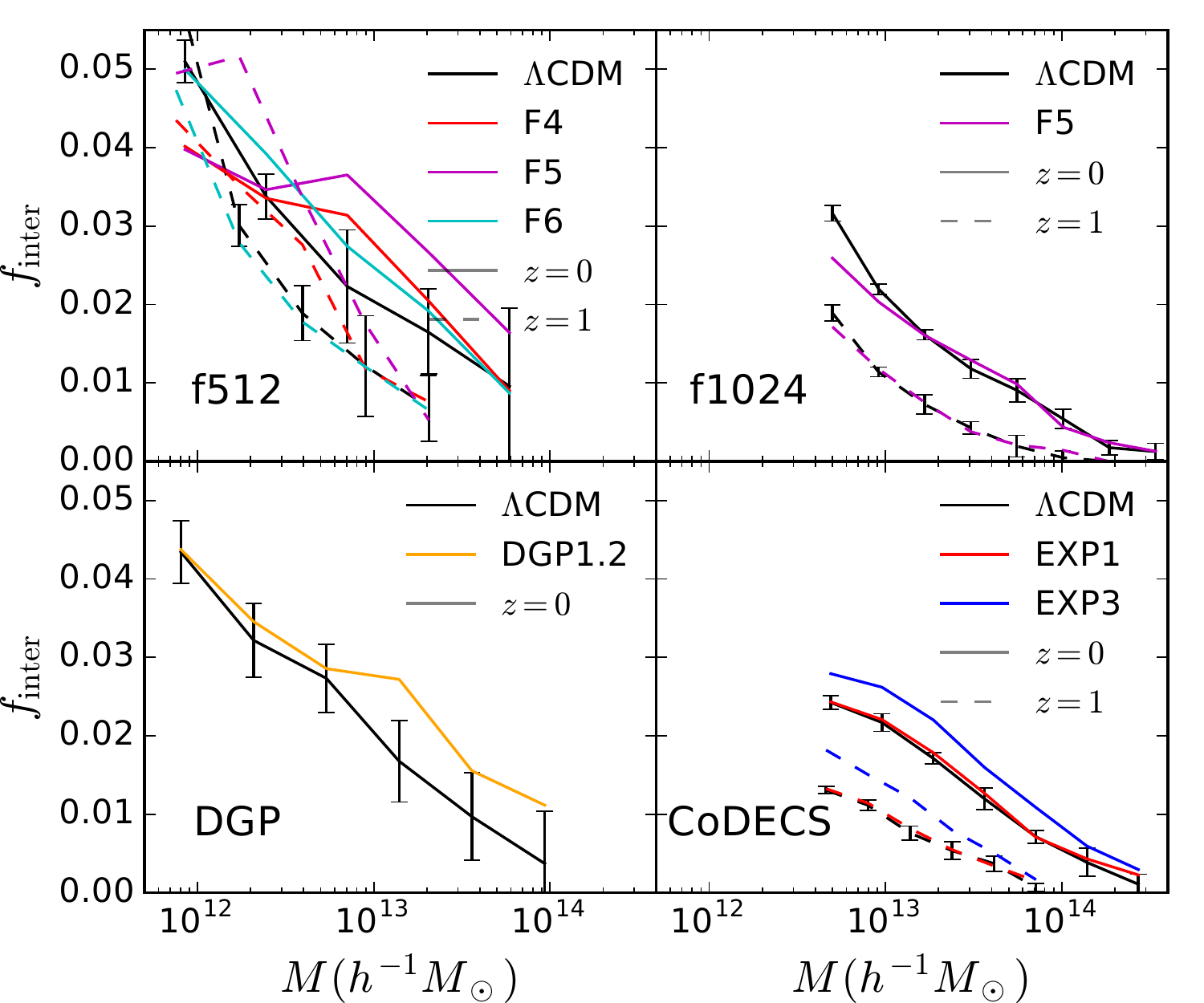}
    \caption{\label{fig:finter_m}
      Interaction rate as a function of mass.
    }
  \end{subfigure}
  \begin{subfigure}{\columnwidth}
    \includegraphics[width=\textwidth]{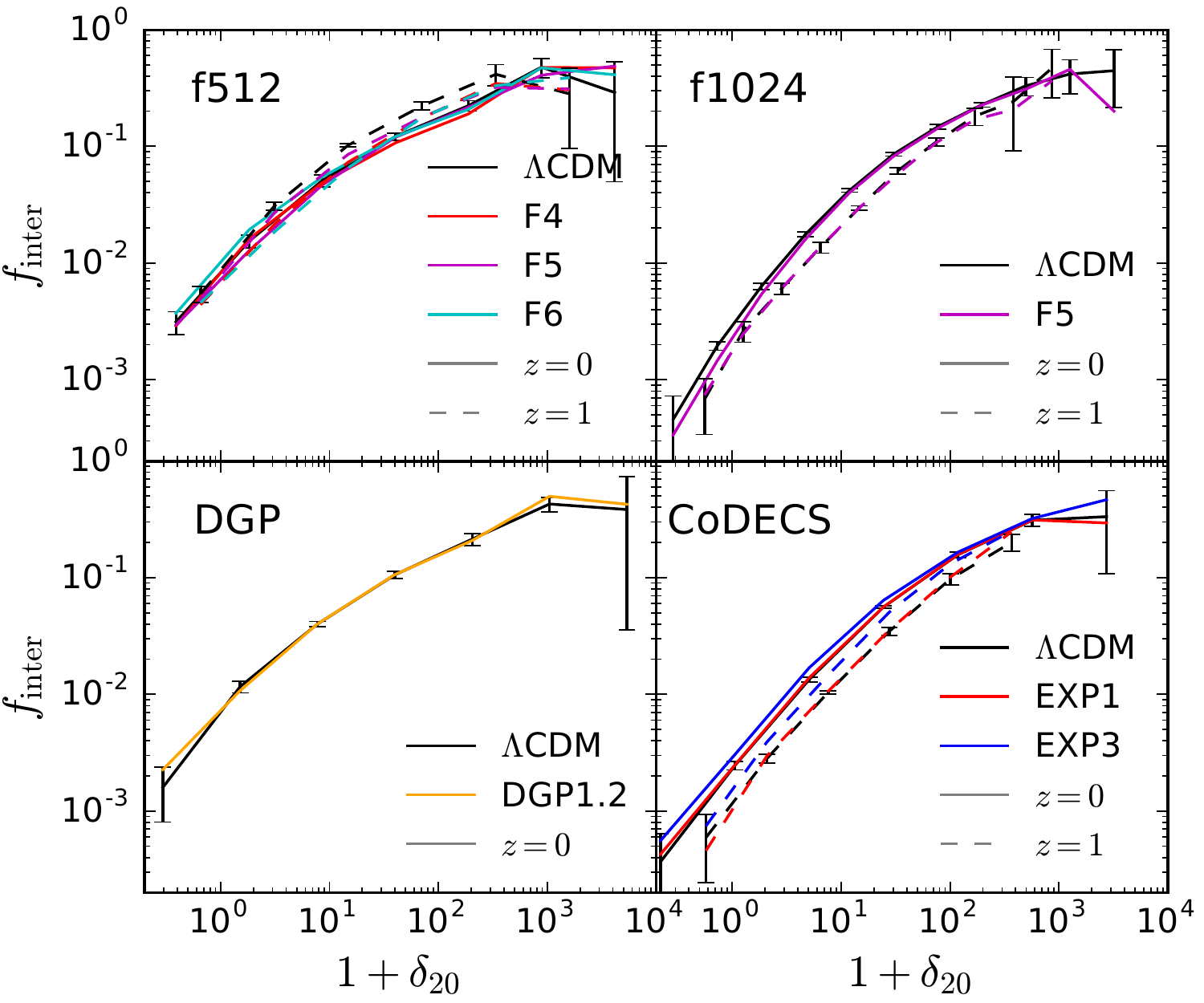}
    \caption{\label{fig:finter_d}%
      Interaction  rate  as a  function  of  density.}
  \end{subfigure}
  \caption{\label{fig:finter}%
    Interaction rate as  a function of mass
    (\subref{fig:finter_m})  and  density (\subref{fig:finter_d})  for
    our different MG and DE models.
  }
\end{figure}

\section{Effects of modified gravity on halo interactions}
\label{sec:inter}

In this section, we study the rate of close interactions
($<2R_\text{vir}$) and their  effects on the alignment of  the spins of
the interacting pair.

\subsection{Interaction rate}

The interaction rate depends on the mass function and the subhalo mass
function,   which    in   turn    depend   on   the    gravity   model.
Fig.~\ref{fig:finter_m}  shows  the  interaction  rates  in  different
gravity models as a function of mass for each set of simulations.
The interaction rate  is defined as the fraction  of targets undergoing
an interaction  with respect  to the  total number  of targets  in the
considered bin.
We divided the simulation volume  in eight equal cubes, and calculated
the mean and standard deviation of the interaction rate in each bin of
mass and density.
For  the sake  of readability,  we only  plot the error-bars for GR, since the other gravity models yield similar uncertainties.

\smallskip

The decreasing shape of the interaction  rate as a function of mass is a consequence of our definition: more massive haloes are less likely to be interacting with a more massive halo \citep{2015MNRAS.451..527L}.
The  f512  and   DGP simulations   have  large  statistical
fluctuations, due to their small  volumes, making it difficult to draw
any conclusion. 
The  other sets  of simulations,  f1024  and CoDECS,  have larger  box
sizes, and thus better statistics. 
In  coupled dark  energy simulations,  the interaction  rate between
EXP1 and GR  are consistent within the error-bars, while  EXP3 has a
higher interaction rate in the whole range.
F5 yields a very similar interaction rate to GR.

Fig.~\ref{fig:finter_d} shows the interaction rate as a function of
the large-scale density $1+\delta_{20}$ for each set of simulation.
Again,  the results  from  small-box simulations  (f512  and DGP)  are
consistent within their large error-bars.
The  tighter  error-bars  in  f1024 show  that  the  interaction  rate
dependence on the large-scale density $\delta_{20}$ is not affected by
$f(R)$ gravity.
In the  coupled dark  energy however,  while EXP1  agrees with  GR, at
all  redshift  and  density,  the   interaction  fraction  in  EXP3  is
systematically higher than in GR for $2\simeq 1+\delta_{20}\simeq 100$
at $z=1$ and $1\simeq 1+\delta_{20}\simeq 20$ at $z=0$.
The excess of interactions is larger at $z=1$ than at $z=0$.
However,  in  this  regime,  the  interaction  fraction  is  very  low
($<0.1$), which makes it difficult to test observationally.

\subsection{%
  Effects of modified gravity on the alignments of interacting pairs
}

In  this section,  we study  the effects  of modified  gravity on  the
alignment of the spins of interacting targets at $z=0$.
In \citet{2017MNRAS.tmp..129L}, we showed that interacting haloes show a
strong alignment, and the strength of the signal increases with mass,
and only weakly depends on density.

\begin{figure*}
  \center
  \includegraphics[width=\textwidth]{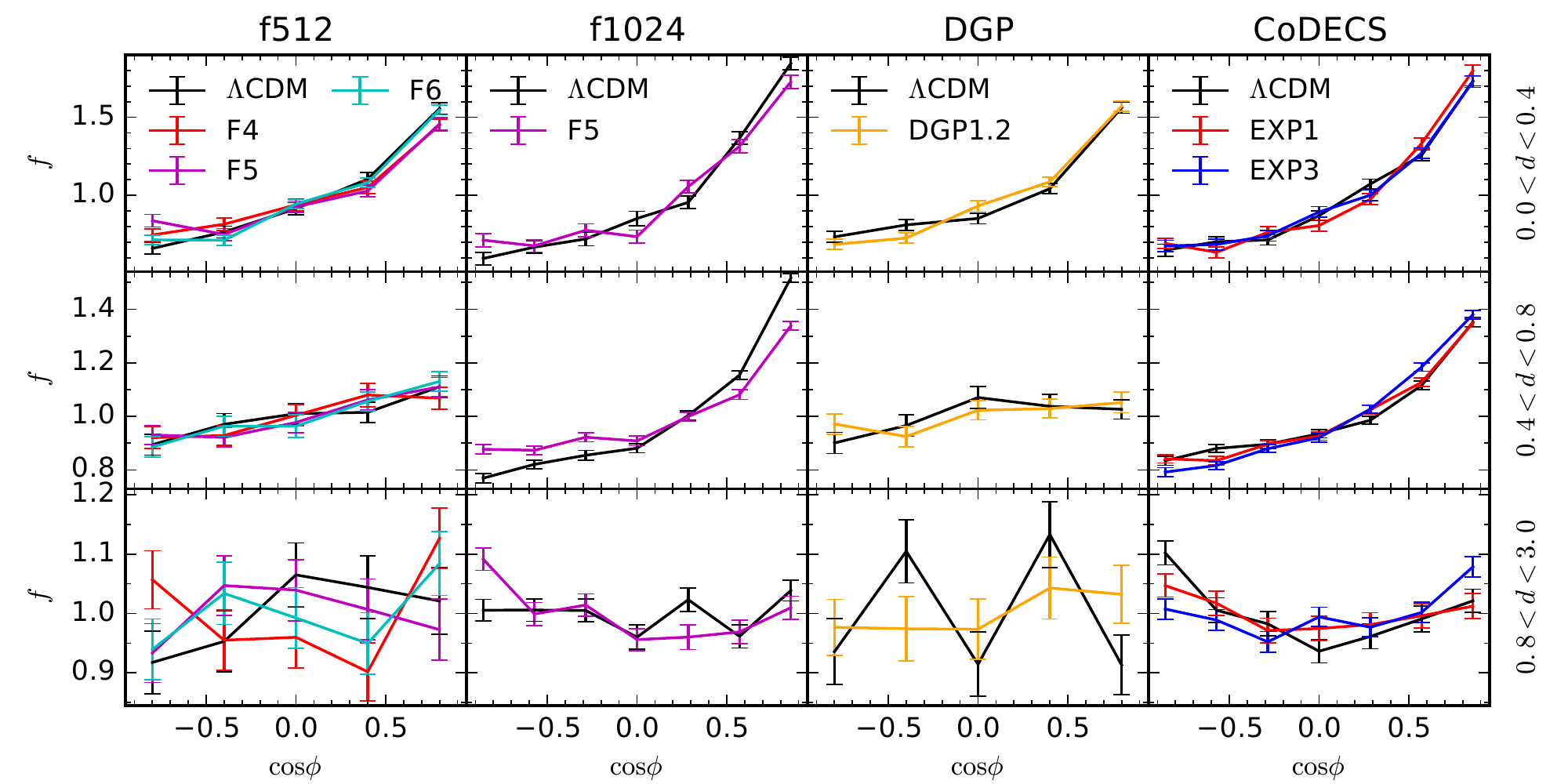}
  \caption{\label{fig:cosphi_r}%
    Normalized pair count of $\cos\phi$, where
    $\phi = (\vect J_\text{T},\vect J_\text{N})$
    as a function of the pair separation $d/d_1$,where $d_1 =  \hMpc{1}$, at $z=0$.
    The pair separation increases from the top to bottom row.
    In the second column, F5 shows a weaker alignment than GR for
    $0.4<d/d_1<0.8$. 
  }
\end{figure*}

\subsubsection{Characterization of the alignment signal}

Following \citet{2005ApJ...628L.101B, 2006MNRAS.369.1293Y,2017MNRAS.tmp..129L},
in order to quantify the alignment of an angle $\theta$ between any
two vectors $(\vect{u}, \vect{v})$,
we proceed as follow.
For a given bin of $\cos\theta$, we count the number of pairs in this
bin $N(\theta)$.
We then randomly shuffle the pairs 100 times, and calculate the mean
and standard deviation $\mean{N(\theta)}$ and $\sigma_\theta$.
We then define the normalized pair count as
\begin{equation}
  \label{eq:npc}
  f (\theta) = \frac{N(\theta)}{\mean{N(\theta)}}.
\end{equation}
The error is then given by $\sigma_\theta / \mean{N(\theta)}$. Random configurations have $f(\theta) = 1$, while alignment have
$f(\cos\theta\simeq \pm 1) \gg 1$, and anti-aligned (orthogonal)
configurations have $f(\cos\theta\simeq 0) \gg 1$.

\subsubsection{Results}

Fig.~\ref{fig:cosphi_r}  shows the  normalized pair  count $f(\phi)$,
where $\phi  = ( \vect{J}_{\text{T}},\vect{J}_\text{N})$, and T and N
respectively denote  the target and  neighbour halo, in the different simulation sets. 
We  divided each  sample into  3 bins  of separation  $0<d<\hMpc{0.4}$
(top), $0.4<d<\hMpc{0.8}$ (middle), and $0.8<d<\hMpc{3}$ (bottom).

In  the  lower   panels,  for  $d>\hMpc{0.8}$,  the  alignment  is
consistent  with  a  random  alignment for  each  model,  showing  no
difference between MG, DE, and GR. 
The  alignment  becomes stronger  as the  pair  separation decreases,  as
expected from stronger tidal forces,  and from previous studies on the
spin correlation function \citep[e.g.,][]{2015MNRAS.450.2195S}. 

In the f512 and DGP simulations, the  small statistics coming
from the small box size do not allow us to see any deviation from GR. 
For f1024 (second column), at  $0.4<d<\hMpc{0.8}$, the alignment signal in F5 is
weaker than in GR (larger excess  of pairs with $\cos\phi \simeq 1$ and lower for $\cos\phi \simeq -1$).
For  $d\leq  \hMpc{0.4}$,  the  alignments  become  consistent  again,
showing almost no deviation from GR.
This can be understood as the action of the fifth force in the F5 model at intermediate separations ($0.4<d<\hMpc{0.8}$), yielding a weaker alignment. 
For small separations, the fifth force is screened and the alignment signal becomes consistent with \lcdm. 
No such behaviour is seen in CoDECS, where the alignment signal is consistent with \lcdm\ for all separations. 
This is consistent with  Fig.~\ref{fig:pql},   where  no  effect   of  DE  on   the  spin distribution could be detected.

\section{Alignment with the large-scale structure}

\label{sec:lss}

\begin{figure*}
  \centering
  \includegraphics[width=\textwidth]{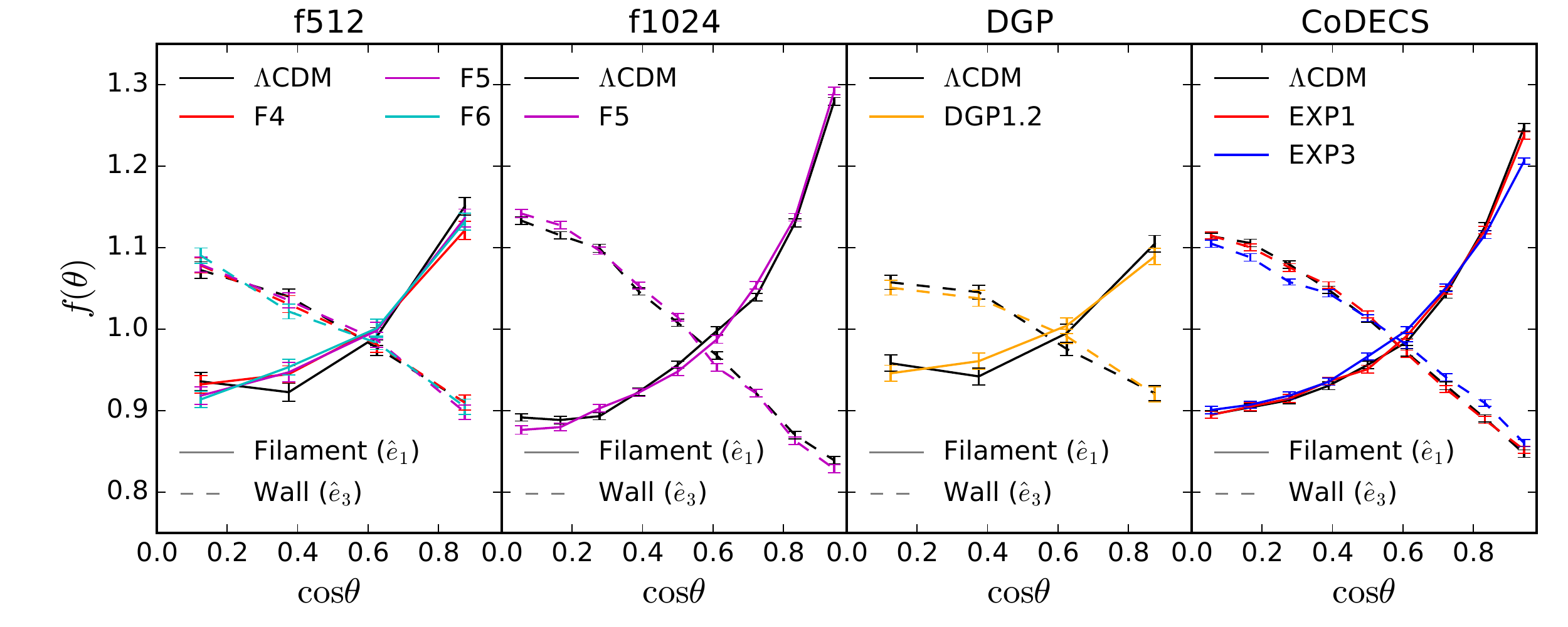}
  \caption{\label{fig:align_lss}%
    Alignment of the major axis with the large-scale structures.
    Shown is the normalized pair count $f(\theta_i)$, where
    $\theta_i = (\vect{a}_\text{T},\hat{e}_i)$ is the angle between
    the major axis of the target and the direction of the wall ($\theta_3$)
    or filament ($\theta_1$).
  }
\end{figure*}

In the  previous section, we  considered the small-scale  alignment of
pairs of interacting haloes.
In this  section, we are  interested in the  alignment of the  spin of
dark matter haloes and the large scale structures, or large-scale alignment.

Alignment of the shapes and spins of haloes with the large-scale
structures have been studied deeply in the literature
\citep[e.g.,][]{2007MNRAS.381...41H,              2009ApJ...706..747Z,
  2012MNRAS.427.3320C,    2013MNRAS.428.2489L,    2013ApJ...762...72T,
  2014MNRAS.443.1090F, 2013ApJ...779..160Z, 2015ApJ...798...17Z}.

There are several ways to characterize the cosmic web.
We     used    the     tidal    tensor     \citep{2007MNRAS.381...41H,
  2009MNRAS.396.1815F}  and  defined   the  large-scale  structure  as
follows.
The  matter density  field  $\delta(\vect  x)$ is  calculated
using a count-in-cell assignment scheme.
The  smooth  density field  $\tilde{\delta}_R$  was  then obtained  by
smoothing  the  matter  density  with  a  Gaussian  kernel  of  radius
$R_\text{G} = \hMpc{4}$.
We then calculated the tidal tensor \tens{T}, defined as
\begin{equation}
  \label{eq:tidal}
  {T}_{ij} = \frac{\partial^2 \tilde\phi}{\partial x_i \partial x_j},
\end{equation}
where  $\tilde\phi$ is  the gravitational  potential, solution  to the
Poisson equation (assuming GR)
\begin{equation}
  \nabla^2\tilde\phi = \tilde{\delta}_R.
\end{equation}
We then calculated the eigenvalues of \tens{T},
$\lambda_1\leq \lambda_2  \leq \lambda_3$,  and defined  voids, walls,
filaments,  and  knots  pixels  with  respectively  0,  1,  2,  and  3
eigenvalues larger than $\lambda_\text{thresh}$.
We set $\lambda_\text{thresh}=0.4$, which visually provided the best
cosmic web.
A discussion about the choice of $\lambda_\text{thresh}$ may be found
in \citet{2009MNRAS.396.1815F,2015MNRAS.447.2683A}.
This calculation was made using the \code{DensTools} code\footnote{The
  code is available at \url{https://github.com/damonge/DensTools}}. 
Each pixel thus has an unambiguously defined environment  (void, wall,
filament, knot). For each halo, we define the environment as that of the pixel where its centre is located.

Fig.~\ref{fig:align_lss} shows the  normalized pair count $f(\theta_i)$
(Eq.~\ref{eq:npc}), where $\theta_i = (\vect a_\text{T},\hat{e}_i)$ is the angle between the major axis of the halo and the direction of the LSS ($\hat e_3$ is the direction normal  to the  walls and  $\hat e_1$  the direction  of the filaments). Since knots and voids do not have a well-defined direction, we exclude them from this analysis.

The  major axis  is  well  aligned with  the  direction of  the
filaments, and  in the  plane of  the wall  (orthogonal to  the normal
direction    of    the    wall),    confirming    previous    findings
\citep[e.g.][]{2009ApJ...706..747Z}. EXP3  shows   the  strongest   deviation  to   GR,  with   weakly  but systematically weaker alignment signal than GR. EXP1 on the other hand shows no deviation at all.

In the f1024  simulations, F5 seems to show a  stronger alignment than
GR, although the significance is weak. A similar trend can be seen in  F256 and DGP, where $f(R)$ models seem to show stronger alignment  than GR, while  DGP seem to show  weaker alignment than GR. However, in  f512 and DGP, the large error-bars do not allow strong conclusions. 


\section{Summary and discussion}

Using cosmological $N$-body simulations with different gravity models,
namely, \lcdm, $f(R)$,  DGP, and coupled dark  energy, we studied the effect of gravity and dark energy on internal properties of subhaloes, namely,  structural and spin parameters. We also studied the effects of  modified gravity on the halo interaction rate.

We then studied the  alignment of
the spins of interacting pairs, following \citet{2017MNRAS.tmp..129L}.
Finally, we  performed for the  first time  a systematic study  of the
alignment  of haloes  with the  large-scale structures  (filaments and
walls) in modified gravity, as well as  of the distribution of the spin parameters in voids and knots. Our findings are summarized below.
\begin{itemize}
\item
  $f(R)$ models  yield a  larger  Bullock spin  parameter, in agreement with 
  \citet{2013ApJ...763...28L} and a lower oblateness and sphericity.
  At $z=0$, the difference in  the structural parameters ($q$ and $s$)
  is smaller than at $z=1$.
\item
  EXP3  yields  larger  oblateness  and  sphericity,  while  the  spin
  parameter us unaffected. 
  EXP1 is essentially indistinguishable from GR.
\item
  The interaction rate is largely unaffected by modified gravity. Strongly coupled dark energy models however show an enhancement of the interaction rate at all masses.  
  \item
  The alignment of the spins  of interacting pairs of haloes decreases
  with the pair  separation for each model. In F5,  the a alignment is
  similar   to  GR   for  $d<\hMpc{0.4}$,   and  becomes   weaker  for
  $0.4<d<\hMpc{0.8}$, due to the effect of the fifth force. Coupled dark energy does not affect the spin alignment  of pairs.  
\item
  The (anti-)alignment of the major  axis of haloes with the direction
  of the filaments (walls) is weaker for the EXP3 model than in GR.
  Simulations with $L\simeq \hMpc{250}$ do not show any departure from
  GR.
\item
 Large volumes ($L\simeq \hGpc{1}$) are needed to distinguish between
 gravity models. 
\end{itemize}

The fact that very large volumes are needed to detect any deviation from GR shows the weakness of the signal. 
For instance, in case of the alignment with the LSS, the large-scale alignment is largely unaffected by  modified gravity. Therefore, one can argue that treatments of intrinsic alignment based on GR should not induce bias in the analysis. 

The strength of our study is to  apply the same method to several sets
of  simulations   with  different   gravity  models,  box   size,  and
resolutions.
This  is the  first  study devoted  to  the study  of  the small-  and
large-scale alignment in modified gravity and dark energy models.

\subsection*{Acknowledgements}
We thank KIAS Center for  Advanced Computation for providing computing
resources. 
HAW is supported by the Beecroft Trust and the European Research Council through grant 646702 (CosTesGrav).
DFM acknowledges support from the Research Council of Norway, and the
NOTUR facilities. 



\bibliographystyle{mnras}
\bibliography{biblio} 

\begin{thebibliography}{}
\makeatletter
\relax
\def\mn@urlcharsother{\let\do\@makeother \do\$\do\&\do\#\do\^\do\_\do\%\do\~}
\def\mn@doi{\begingroup\mn@urlcharsother \@ifnextchar [ {\mn@doi@}
  {\mn@doi@[]}}
\def\mn@doi@[#1]#2{\def\@tempa{#1}\ifx\@tempa\@empty \href
  {http://dx.doi.org/#2} {doi:#2}\else \href {http://dx.doi.org/#2} {#1}\fi
  \endgroup}
\def\mn@eprint#1#2{\mn@eprint@#1:#2::\@nil}
\def\mn@eprint@arXiv#1{\href {http://arxiv.org/abs/#1} {{\tt arXiv:#1}}}
\def\mn@eprint@dblp#1{\href {http://dblp.uni-trier.de/rec/bibtex/#1.xml}
  {dblp:#1}}
\def\mn@eprint@#1:#2:#3:#4\@nil{\def\@tempa {#1}\def\@tempb {#2}\def\@tempc
  {#3}\ifx \@tempc \@empty \let \@tempc \@tempb \let \@tempb \@tempa \fi \ifx
  \@tempb \@empty \def\@tempb {arXiv}\fi \@ifundefined
  {mn@eprint@\@tempb}{\@tempb:\@tempc}{\expandafter \expandafter \csname
  mn@eprint@\@tempb\endcsname \expandafter{\@tempc}}}

\bibitem[\protect\citeauthoryear{{Alonso}, {Eardley}  \& {Peacock}}{{Alonso}
  et~al.}{2015}]{2015MNRAS.447.2683A}
{Alonso} D.,  {Eardley} E.,   {Peacock} J.~A.,  2015, \mn@doi [\mnras]
  {10.1093/mnras/stu2632}, \href
  {http://adsabs.harvard.edu/abs/2015MNRAS.447.2683A} {447, 2683}

\bibitem[\protect\citeauthoryear{{Amendola}}{{Amendola}}{2000}]{2000PhRvD..62d3511A}
{Amendola} L.,  2000, \mn@doi [\prd] {10.1103/PhysRevD.62.043511}, \href
  {http://adsabs.harvard.edu/abs/2000PhRvD..62d3511A} {62, 043511}

\bibitem[\protect\citeauthoryear{Amendola \& Group}{Amendola \&
  Group}{2013}]{lrr-2013-6}
Amendola L.,  Group E. T.~W.,  2013, \mn@doi [Living Reviews in Relativity]
  {10.1007/lrr-2013-6}, 16

\bibitem[\protect\citeauthoryear{{Baldi}}{{Baldi}}{2012}]{2012MNRAS.422.1028B}
{Baldi} M.,  2012, \mn@doi [\mnras] {10.1111/j.1365-2966.2012.20675.x}, \href
  {http://adsabs.harvard.edu/abs/2012MNRAS.422.1028B} {422, 1028}

\bibitem[\protect\citeauthoryear{{Baldi}, {Pettorino}, {Robbers}  \&
  {Springel}}{{Baldi} et~al.}{2010}]{2010MNRAS.403.1684B}
{Baldi} M.,  {Pettorino} V.,  {Robbers} G.,   {Springel} V.,  2010, \mn@doi
  [\mnras] {10.1111/j.1365-2966.2009.15987.x}, \href
  {http://adsabs.harvard.edu/abs/2010MNRAS.403.1684B} {403, 1684}

\bibitem[\protect\citeauthoryear{{Brainerd}}{{Brainerd}}{2005}]{2005ApJ...628L.101B}
{Brainerd} T.~G.,  2005, \mn@doi [\apjl] {10.1086/432713}, \href
  {http://adsabs.harvard.edu/abs/2005ApJ...628L.101B} {628, L101}

\bibitem[\protect\citeauthoryear{{Brax}, {van de Bruck}, {Davis}  \&
  {Shaw}}{{Brax} et~al.}{2008}]{2008PhRvD..78j4021B}
{Brax} P.,  {van de Bruck} C.,  {Davis} A.-C.,   {Shaw} D.~J.,  2008, \mn@doi
  [\prd] {10.1103/PhysRevD.78.104021}, \href
  {http://adsabs.harvard.edu/abs/2008PhRvD..78j4021B} {78, 104021}

\bibitem[\protect\citeauthoryear{{Bryan} \& {Norman}}{{Bryan} \&
  {Norman}}{1998}]{1998ApJ...495...80B}
{Bryan} G.~L.,  {Norman} M.~L.,  1998, \mn@doi [\apj] {10.1086/305262}, \href
  {http://adsabs.harvard.edu/abs/1998ApJ...495...80B} {495, 80}

\bibitem[\protect\citeauthoryear{{Bullock}, {Dekel}, {Kolatt}, {Kravtsov},
  {Klypin}, {Porciani}  \& {Primack}}{{Bullock}
  et~al.}{2001}]{2001ApJ...555..240B}
{Bullock} J.~S.,  {Dekel} A.,  {Kolatt} T.~S.,  {Kravtsov} A.~V.,  {Klypin}
  A.~A.,  {Porciani} C.,   {Primack} J.~R.,  2001, \mn@doi [\apj]
  {10.1086/321477}, \href {http://adsabs.harvard.edu/abs/2001ApJ...555..240B}
  {555, 240}

\bibitem[\protect\citeauthoryear{{Casas}, {Amendola}, {Baldi}, {Pettorino}  \&
  {Vollmer}}{{Casas} et~al.}{2016}]{2016JCAP...01..045C}
{Casas} S.,  {Amendola} L.,  {Baldi} M.,  {Pettorino} V.,   {Vollmer} A.,
  2016, \mn@doi [\jcap] {10.1088/1475-7516/2016/01/045}, \href
  {http://adsabs.harvard.edu/abs/2016JCAP...01..045C} {1, 045}

\bibitem[\protect\citeauthoryear{{Clifton}, {Ferreira}, {Padilla}  \&
  {Skordis}}{{Clifton} et~al.}{2012}]{2012PhR...513....1C}
{Clifton} T.,  {Ferreira} P.~G.,  {Padilla} A.,   {Skordis} C.,  2012, \mn@doi
  [\physrep] {10.1016/j.physrep.2012.01.001}, \href
  {http://adsabs.harvard.edu/abs/2012PhR...513....1C} {513, 1}

\bibitem[\protect\citeauthoryear{{Codis}, {Pichon}, {Devriendt}, {Slyz},
  {Pogosyan}, {Dubois}  \& {Sousbie}}{{Codis}
  et~al.}{2012}]{2012MNRAS.427.3320C}
{Codis} S.,  {Pichon} C.,  {Devriendt} J.,  {Slyz} A.,  {Pogosyan} D.,
  {Dubois} Y.,   {Sousbie} T.,  2012, \mn@doi [\mnras]
  {10.1111/j.1365-2966.2012.21636.x}, \href
  {http://adsabs.harvard.edu/abs/2012MNRAS.427.3320C} {427, 3320}

\bibitem[\protect\citeauthoryear{{Colless}}{{Colless}}{1999}]{1999RSPTA.357..105C}
{Colless} M.,  1999, \mn@doi [Royal Society of London Philosophical
  Transactions Series A] {10.1098/rsta.1999.0317}, \href
  {http://adsabs.harvard.edu/abs/1999RSPTA.357..105C} {357, 105}

\bibitem[\protect\citeauthoryear{{Copeland}, {Sami}  \& {Tsujikawa}}{{Copeland}
  et~al.}{2006}]{2006IJMPD..15.1753C}
{Copeland} E.~J.,  {Sami} M.,   {Tsujikawa} S.,  2006, \mn@doi [International
  Journal of Modern Physics D] {10.1142/S021827180600942X}, \href
  {http://adsabs.harvard.edu/abs/2006IJMPD..15.1753C} {15, 1753}

\bibitem[\protect\citeauthoryear{{Dvali}, {Gabadadze}  \& {Porrati}}{{Dvali}
  et~al.}{2000}]{2000PhLB..485..208D}
{Dvali} G.,  {Gabadadze} G.,   {Porrati} M.,  2000, \mn@doi [Physics Letters B]
  {10.1016/S0370-2693(00)00669-9}, \href
  {http://adsabs.harvard.edu/abs/2000PhLB..485..208D} {485, 208}

\bibitem[\protect\citeauthoryear{{Forero-Romero}, {Hoffman}, {Gottl{\"o}ber},
  {Klypin}  \& {Yepes}}{{Forero-Romero} et~al.}{2009}]{2009MNRAS.396.1815F}
{Forero-Romero} J.~E.,  {Hoffman} Y.,  {Gottl{\"o}ber} S.,  {Klypin} A.,
  {Yepes} G.,  2009, \mn@doi [\mnras] {10.1111/j.1365-2966.2009.14885.x}, \href
  {http://adsabs.harvard.edu/abs/2009MNRAS.396.1815F} {396, 1815}

\bibitem[\protect\citeauthoryear{{Forero-Romero}, {Contreras}  \&
  {Padilla}}{{Forero-Romero} et~al.}{2014}]{2014MNRAS.443.1090F}
{Forero-Romero} J.~E.,  {Contreras} S.,   {Padilla} N.,  2014, \mn@doi [\mnras]
  {10.1093/mnras/stu1150}, \href
  {http://adsabs.harvard.edu/abs/2014MNRAS.443.1090F} {443, 1090}

\bibitem[\protect\citeauthoryear{Gannouji, Moraes, Mota, Polarski, Tsujikawa
  \& Winther}{Gannouji et~al.}{2010}]{Gannouji:2010fc}
Gannouji R.,  Moraes B.,  Mota D.~F.,  Polarski D.,  Tsujikawa S.,   Winther
  H.~A.,  2010, \mn@doi [Phys. Rev.] {10.1103/PhysRevD.82.124006}, D82, 124006

\bibitem[\protect\citeauthoryear{{Gill}, {Knebe}  \& {Gibson}}{{Gill}
  et~al.}{2004}]{2004MNRAS.351..399G}
{Gill} S.~P.~D.,  {Knebe} A.,   {Gibson} B.~K.,  2004, \mn@doi [\mnras]
  {10.1111/j.1365-2966.2004.07786.x}, \href
  {http://adsabs.harvard.edu/abs/2004MNRAS.351..399G} {351, 399}

\bibitem[\protect\citeauthoryear{{Hahn}, {Carollo}, {Porciani}  \&
  {Dekel}}{{Hahn} et~al.}{2007}]{2007MNRAS.381...41H}
{Hahn} O.,  {Carollo} C.~M.,  {Porciani} C.,   {Dekel} A.,  2007, \mn@doi
  [\mnras] {10.1111/j.1365-2966.2007.12249.x}, \href
  {http://adsabs.harvard.edu/abs/2007MNRAS.381...41H} {381, 41}

\bibitem[\protect\citeauthoryear{{Hu} \& {Sawicki}}{{Hu} \&
  {Sawicki}}{2007}]{2007PhRvD..76f4004H}
{Hu} W.,  {Sawicki} I.,  2007, \mn@doi [\prd] {10.1103/PhysRevD.76.064004},
  \href {http://adsabs.harvard.edu/abs/2007PhRvD..76f4004H} {76, 064004}

\bibitem[\protect\citeauthoryear{Joyce, Jain, Khoury  \& Trodden}{Joyce
  et~al.}{2015}]{joyce}
Joyce A.,  Jain B.,  Khoury J.,   Trodden M.,  2015, \mn@doi [Phys. Rept.]
  {10.1016/j.physrep.2014.12.002}, 568, 1

\bibitem[\protect\citeauthoryear{Khoury}{Khoury}{2010}]{Khoury2010}
Khoury J.,  2010, preprint (\mn@eprint {} {1011.5909})

\bibitem[\protect\citeauthoryear{{Kim} \& {Park}}{{Kim} \&
  {Park}}{2006}]{2006ApJ...639..600K}
{Kim} J.,  {Park} C.,  2006, \mn@doi [\apj] {10.1086/499761}, \href
  {http://adsabs.harvard.edu/abs/2006ApJ...639..600K} {639, 600}

\bibitem[\protect\citeauthoryear{{Kim}, {Park}, {L'Huillier}  \& {Hong}}{{Kim}
  et~al.}{2015}]{2015JKAS...48..213K}
{Kim} J.,  {Park} C.,  {L'Huillier} B.,   {Hong} S.~E.,  2015, \mn@doi [J.
  Korean Astron. Soc.] {10.5303/JKAS.2015.48.4.213}, \href
  {http://adsabs.harvard.edu/abs/2015JKAS...48..213K} {48, 213}

\bibitem[\protect\citeauthoryear{{Knollmann} \& {Knebe}}{{Knollmann} \&
  {Knebe}}{2009}]{2009ApJS..182..608K}
{Knollmann} S.~R.,  {Knebe} A.,  2009, \mn@doi [\apjs]
  {10.1088/0067-0049/182/2/608}, \href
  {http://cdsads.u-strasbg.fr/abs/2009ApJS..182..608K} {182, 608}

\bibitem[\protect\citeauthoryear{Koivisto}{Koivisto}{2005}]{tomi}
Koivisto T.,  2005, \mn@doi [Phys. Rev.] {10.1103/PhysRevD.72.043516}, D72,
  043516

\bibitem[\protect\citeauthoryear{Koivisto \& Mota}{Koivisto \&
  Mota}{2007}]{koivisto}
Koivisto T.,  Mota D.~F.,  2007, \mn@doi [Phys. Lett.]
  {10.1016/j.physletb.2006.11.048}, B644, 104

\bibitem[\protect\citeauthoryear{{L'Huillier}, {Park}  \& {Kim}}{{L'Huillier}
  et~al.}{2015}]{2015MNRAS.451..527L}
{L'Huillier} B.,  {Park} C.,   {Kim} J.,  2015, \mn@doi [\mnras]
  {10.1093/mnras/stv995}, \href
  {http://adsabs.harvard.edu/abs/2015MNRAS.451..527L} {451, 527}

\bibitem[\protect\citeauthoryear{{L'Huillier}, {Park}  \& {Kim}}{{L'Huillier}
  et~al.}{2017}]{2017MNRAS.tmp..129L}
{L'Huillier} B.,  {Park} C.,   {Kim} J.,  2017, preprint, \href
  {http://adsabs.harvard.edu/abs/2017arXiv170104417L} {} (\mn@eprint {arXiv}
  {1701.04417})

\bibitem[\protect\citeauthoryear{{Laureijs} et~al.,}{{Laureijs}
  et~al.}{2011}]{2011arXiv1110.3193L}
{Laureijs} R.,  et~al., 2011, preprint, \href
  {http://adsabs.harvard.edu/abs/2011arXiv1110.3193L} {} (\mn@eprint {arXiv}
  {1110.3193})

\bibitem[\protect\citeauthoryear{{Lee}, {Zhao}, {Li}  \& {Koyama}}{{Lee}
  et~al.}{2013}]{2013ApJ...763...28L}
{Lee} J.,  {Zhao} G.-B.,  {Li} B.,   {Koyama} K.,  2013, \mn@doi [\apj]
  {10.1088/0004-637X/763/1/28}, \href
  {http://adsabs.harvard.edu/abs/2013ApJ...763...28L} {763, 28}

\bibitem[\protect\citeauthoryear{Leithes, Malik, Mulryne  \& Nunes}{Leithes
  et~al.}{2016}]{lei}
Leithes A.,  Malik K.~A.,  Mulryne D.~J.,   Nunes N.~J.,  2016

\bibitem[\protect\citeauthoryear{{Li}}{{Li}}{2011}]{Li:2011c}
{Li} B.,  2011, \mn@doi [\mnras] {10.1111/j.1365-2966.2010.17867.x}, \href
  {http://adsabs.harvard.edu/abs/2011MNRAS.411.2615L} {411, 2615}

\bibitem[\protect\citeauthoryear{{Li} \& {Zhao}}{{Li} \&
  {Zhao}}{2010}]{2010PhRvD..81j4047L}
{Li} B.,  {Zhao} H.,  2010, \mn@doi [\prd] {10.1103/PhysRevD.81.104047}, \href
  {http://adsabs.harvard.edu/abs/2010PhRvD..81j4047L} {81, 104047}

\bibitem[\protect\citeauthoryear{{Li}, {Mota}  \& {Barrow}}{{Li}
  et~al.}{2011}]{Li:2011b}
{Li} B.,  {Mota} D.~F.,   {Barrow} J.~D.,  2011, \mn@doi [\apj]
  {10.1088/0004-637X/728/2/109}, \href
  {http://adsabs.harvard.edu/abs/2011ApJ...728..109L} {728, 109}

\bibitem[\protect\citeauthoryear{{Li}, {Zhao}  \& {Koyama}}{{Li}
  et~al.}{2012}]{2012MNRAS.421.3481L}
{Li} B.,  {Zhao} G.-B.,   {Koyama} K.,  2012, \mn@doi [\mnras]
  {10.1111/j.1365-2966.2012.20573.x}, \href
  {http://adsabs.harvard.edu/abs/2012MNRAS.421.3481L} {421, 3481}

\bibitem[\protect\citeauthoryear{{Libeskind}, {Hoffman}, {Forero-Romero},
  {Gottl{\"o}ber}, {Knebe}, {Steinmetz}  \& {Klypin}}{{Libeskind}
  et~al.}{2013a}]{2013MNRAS.428.2489L}
{Libeskind} N.~I.,  {Hoffman} Y.,  {Forero-Romero} J.,  {Gottl{\"o}ber} S.,
  {Knebe} A.,  {Steinmetz} M.,   {Klypin} A.,  2013a, \mn@doi [\mnras]
  {10.1093/mnras/sts216}, \href
  {http://adsabs.harvard.edu/abs/2013MNRAS.428.2489L} {428, 2489}

\bibitem[\protect\citeauthoryear{{Libeskind}, {Hoffman}, {Steinmetz},
  {Gottl{\"o}ber}, {Knebe}  \& {Hess}}{{Libeskind}
  et~al.}{2013b}]{2013ApJ...766L..15L}
{Libeskind} N.~I.,  {Hoffman} Y.,  {Steinmetz} M.,  {Gottl{\"o}ber} S.,
  {Knebe} A.,   {Hess} S.,  2013b, \mn@doi [\apjl]
  {10.1088/2041-8205/766/2/L15}, \href
  {http://adsabs.harvard.edu/abs/2013ApJ...766L..15L} {766, L15}

\bibitem[\protect\citeauthoryear{{Llinares} \& {Mota}}{{Llinares} \&
  {Mota}}{2013a}]{Llinares:2013}
{Llinares} C.,  {Mota} D.~F.,  2013a, \mn@doi [Physical Review Letters]
  {10.1103/PhysRevLett.110.151104}, \href
  {http://adsabs.harvard.edu/abs/2013PhRvL.110o1104L} {110, 151104}

\bibitem[\protect\citeauthoryear{Llinares \& Mota}{Llinares \&
  Mota}{2013b}]{Llinares:2013qbh}
Llinares C.,  Mota D.,  2013b, \mn@doi [Phys. Rev. Lett.]
  {10.1103/PhysRevLett.110.161101}, 110, 161101

\bibitem[\protect\citeauthoryear{Llinares, Mota  \& Winther}{Llinares
  et~al.}{2014a}]{isis}
Llinares C.,  Mota D.~F.,   Winther H.~A.,  2014a, \mn@doi [Astron. Astrophys.]
  {10.1051/0004-6361/201322412}, 562, A78

\bibitem[\protect\citeauthoryear{{Llinares}, {Mota}  \& {Winther}}{{Llinares}
  et~al.}{2014b}]{2014A&A...562A..78L}
{Llinares} C.,  {Mota} D.~F.,   {Winther} H.~A.,  2014b, \mn@doi [\aap]
  {10.1051/0004-6361/201322412}, \href
  {http://adsabs.harvard.edu/abs/2014A\%26A...562A..78L} {562, A78}

\bibitem[\protect\citeauthoryear{Maartens \& Koyama}{Maartens \&
  Koyama}{2010}]{lrr-2010-5}
Maartens R.,  Koyama K.,  2010, \mn@doi [Living Reviews in Relativity]
  {10.1007/lrr-2010-5}, 13

\bibitem[\protect\citeauthoryear{{Monaghan} \& {Lattanzio}}{{Monaghan} \&
  {Lattanzio}}{1985}]{1985A&A...149..135M}
{Monaghan} J.~J.,  {Lattanzio} J.~C.,  1985, \aap, \href
  {http://adsabs.harvard.edu/abs/1985A\%26A...149..135M} {149, 135}

\bibitem[\protect\citeauthoryear{Mota, Kristiansen, Koivisto  \&
  Groeneboom}{Mota et~al.}{2007}]{Mota:2007sz}
Mota D.~F.,  Kristiansen J.~R.,  Koivisto T.,   Groeneboom N.~E.,  2007,
  \mn@doi [Mon. Not. Roy. Astron. Soc.] {10.1111/j.1365-2966.2007.12413.x},
  382, 793

\bibitem[\protect\citeauthoryear{Mota, Shaw  \& Silk}{Mota
  et~al.}{2008}]{Mota:2007zn}
Mota D.~F.,  Shaw D.~J.,   Silk J.,  2008, \mn@doi [Astrophys. J.]
  {10.1086/524401}, 675, 29

\bibitem[\protect\citeauthoryear{Mota, Sandstad  \& Zlosnik}{Mota
  et~al.}{2010}]{sandstad}
Mota D.~F.,  Sandstad M.,   Zlosnik T.,  2010, \mn@doi [JHEP]
  {10.1007/JHEP12(2010)051}, 12, 051

\bibitem[\protect\citeauthoryear{{Peebles}}{{Peebles}}{1969}]{1969ApJ...155..393P}
{Peebles} P.~J.~E.,  1969, \mn@doi [\apj] {10.1086/149876}, \href
  {http://adsabs.harvard.edu/abs/1969ApJ...155..393P} {155, 393}

\bibitem[\protect\citeauthoryear{{Perlmutter} et~al.,}{{Perlmutter}
  et~al.}{1998}]{1998Natur.391...51P}
{Perlmutter} S.,  et~al., 1998, \mn@doi [\nat] {10.1038/34124}, \href
  {http://adsabs.harvard.edu/abs/1998Natur.391...51P} {391, 51}

\bibitem[\protect\citeauthoryear{{Pettorino}}{{Pettorino}}{2013}]{Pettorino:2013}
{Pettorino} V.,  2013, \mn@doi [\prd] {10.1103/PhysRevD.88.063519}, \href
  {http://cdsads.u-strasbg.fr/abs/2013PhRvD..88f3519P} {88, 063519}

\bibitem[\protect\citeauthoryear{{Riess} et~al.,}{{Riess}
  et~al.}{1998}]{1998AJ....116.1009R}
{Riess} A.~G.,  et~al., 1998, \mn@doi [\aj] {10.1086/300499}, \href
  {http://adsabs.harvard.edu/abs/1998AJ....116.1009R} {116, 1009}

\bibitem[\protect\citeauthoryear{{Sabiu}, {Mota}, {Llinares}  \&
  {Park}}{{Sabiu} et~al.}{2016}]{2016A&A...592A..38S}
{Sabiu} C.~G.,  {Mota} D.~F.,  {Llinares} C.,   {Park} C.,  2016, \mn@doi
  [\aap] {10.1051/0004-6361/201527776}, \href
  {http://adsabs.harvard.edu/abs/2016A%26A...592A..38S} {592, A38}

\bibitem[\protect\citeauthoryear{{Shi}, {Li}, {Han}, {Gao}  \&
  {Hellwing}}{{Shi} et~al.}{2015}]{2015MNRAS.452.3179S}
{Shi} D.,  {Li} B.,  {Han} J.,  {Gao} L.,   {Hellwing} W.~A.,  2015, \mn@doi
  [\mnras] {10.1093/mnras/stv1549}, \href
  {http://adsabs.harvard.edu/abs/2015MNRAS.452.3179S} {452, 3179}

\bibitem[\protect\citeauthoryear{{Singh}, {Mandelbaum}  \& {More}}{{Singh}
  et~al.}{2015}]{2015MNRAS.450.2195S}
{Singh} S.,  {Mandelbaum} R.,   {More} S.,  2015, \mn@doi [\mnras]
  {10.1093/mnras/stv778}, \href
  {http://adsabs.harvard.edu/abs/2015MNRAS.450.2195S} {450, 2195}

\bibitem[\protect\citeauthoryear{{Springel}}{{Springel}}{2005}]{2005MNRAS.364.1105S}
{Springel} V.,  2005, \mn@doi [\mnras] {10.1111/j.1365-2966.2005.09655.x},
  \href {http://adsabs.harvard.edu/abs/2005MNRAS.364.1105S} {364, 1105}

\bibitem[\protect\citeauthoryear{{Springel}, {White}, {Tormen}  \&
  {Kauffmann}}{{Springel} et~al.}{2001}]{2001MNRAS.328..726S}
{Springel} V.,  {White} S.~D.~M.,  {Tormen} G.,   {Kauffmann} G.,  2001,
  \mn@doi [\mnras] {10.1046/j.1365-8711.2001.04912.x}, \href
  {http://adsabs.harvard.edu/abs/2001MNRAS.328..726S} {328, 726}

\bibitem[\protect\citeauthoryear{{Teyssier}}{{Teyssier}}{2002}]{2002A&A...385..337T}
{Teyssier} R.,  2002, \mn@doi [\aap] {10.1051/0004-6361:20011817}, \href
  {http://adsabs.harvard.edu/abs/2002A%26A...385..337T} {385, 337}

\bibitem[\protect\citeauthoryear{{Trowland}, {Lewis}  \&
  {Bland-Hawthorn}}{{Trowland} et~al.}{2013}]{2013ApJ...762...72T}
{Trowland} H.~E.,  {Lewis} G.~F.,   {Bland-Hawthorn} J.,  2013, \mn@doi [\apj]
  {10.1088/0004-637X/762/2/72}, \href
  {http://adsabs.harvard.edu/abs/2013ApJ...762...72T} {762, 72}

\bibitem[\protect\citeauthoryear{{Wetterich}}{{Wetterich}}{1995}]{1995A&A...301..321W}
{Wetterich} C.,  1995, \aap, \href
  {http://adsabs.harvard.edu/abs/1995A\%26A...301..321W} {301, 321}

\bibitem[\protect\citeauthoryear{{Will}}{{Will}}{2006}]{Will2006LRR.....9....3W}
{Will} C.~M.,  2006, \mn@doi [Living Reviews in Relativity]
  {10.12942/lrr-2006-3}, \href
  {http://adsabs.harvard.edu/abs/2006LRR.....9....3W} {9, 3}

\bibitem[\protect\citeauthoryear{{Will}}{{Will}}{2014}]{2014LRR....17....4W}
{Will} C.~M.,  2014, \mn@doi [Living Reviews in Relativity]
  {10.12942/lrr-2014-4}, \href
  {http://adsabs.harvard.edu/abs/2014LRR....17....4W} {17, 4}

\bibitem[\protect\citeauthoryear{{Winther} et~al.,}{{Winther}
  et~al.}{2015}]{2015MNRAS.454.4208W}
{Winther} H.~A.,  et~al., 2015, \mn@doi [\mnras] {10.1093/mnras/stv2253}, \href
  {http://adsabs.harvard.edu/abs/2015MNRAS.454.4208W} {454, 4208}

\bibitem[\protect\citeauthoryear{{Yang}, {van den Bosch}, {Mo}, {Mao}, {Kang},
  {Weinmann}, {Guo}  \& {Jing}}{{Yang} et~al.}{2006}]{2006MNRAS.369.1293Y}
{Yang} X.,  {van den Bosch} F.~C.,  {Mo} H.~J.,  {Mao} S.,  {Kang} X.,
  {Weinmann} S.~M.,  {Guo} Y.,   {Jing} Y.~P.,  2006, \mn@doi [\mnras]
  {10.1111/j.1365-2966.2006.10373.x}, \href
  {http://adsabs.harvard.edu/abs/2006MNRAS.369.1293Y} {369, 1293}

\bibitem[\protect\citeauthoryear{{Zhang}, {Yang}, {Faltenbacher}, {Springel},
  {Lin}  \& {Wang}}{{Zhang} et~al.}{2009}]{2009ApJ...706..747Z}
{Zhang} Y.,  {Yang} X.,  {Faltenbacher} A.,  {Springel} V.,  {Lin} W.,   {Wang}
  H.,  2009, \mn@doi [\apj] {10.1088/0004-637X/706/1/747}, \href
  {http://adsabs.harvard.edu/abs/2009ApJ...706..747Z} {706, 747}

\bibitem[\protect\citeauthoryear{{Zhang}, {Yang}, {Wang}, {Wang}, {Mo}  \& {van
  den Bosch}}{{Zhang} et~al.}{2013}]{2013ApJ...779..160Z}
{Zhang} Y.,  {Yang} X.,  {Wang} H.,  {Wang} L.,  {Mo} H.~J.,   {van den Bosch}
  F.~C.,  2013, \mn@doi [\apj] {10.1088/0004-637X/779/2/160}, \href
  {http://adsabs.harvard.edu/abs/2013ApJ...779..160Z} {779, 160}

\bibitem[\protect\citeauthoryear{{Zhang}, {Yang}, {Wang}, {Wang}, {Luo}, {Mo}
  \& {van den Bosch}}{{Zhang} et~al.}{2015}]{2015ApJ...798...17Z}
{Zhang} Y.,  {Yang} X.,  {Wang} H.,  {Wang} L.,  {Luo} W.,  {Mo} H.~J.,   {van
  den Bosch} F.~C.,  2015, \mn@doi [\apj] {10.1088/0004-637X/798/1/17}, \href
  {http://adsabs.harvard.edu/abs/2015ApJ...798...17Z} {798, 17}

\bibitem[\protect\citeauthoryear{{Zhao}, {Li}  \& {Koyama}}{{Zhao}
  et~al.}{2011}]{2011PhRvL.107g1303Z}
{Zhao} G.-B.,  {Li} B.,   {Koyama} K.,  2011, \mn@doi [Physical Review Letters]
  {10.1103/PhysRevLett.107.071303}, \href
  {http://adsabs.harvard.edu/abs/2011PhRvL.107g1303Z} {107, 071303}

\bibitem[\protect\citeauthoryear{{de Felice} \& {Tsujikawa}}{{de Felice} \&
  {Tsujikawa}}{2010}]{2010LRR....13....3D}
{de Felice} A.,  {Tsujikawa} S.,  2010, \mn@doi [Living Reviews in Relativity]
  {10.12942/lrr-2010-3}, \href
  {http://adsabs.harvard.edu/abs/2010LRR....13....3D} {13}

\makeatother
\end{thebibliography}






\appendix
\label{sec:spin_pb}

\section{Peebles versus Bullock spins}

\begin{figure}
  \includegraphics[width=.45\textwidth]{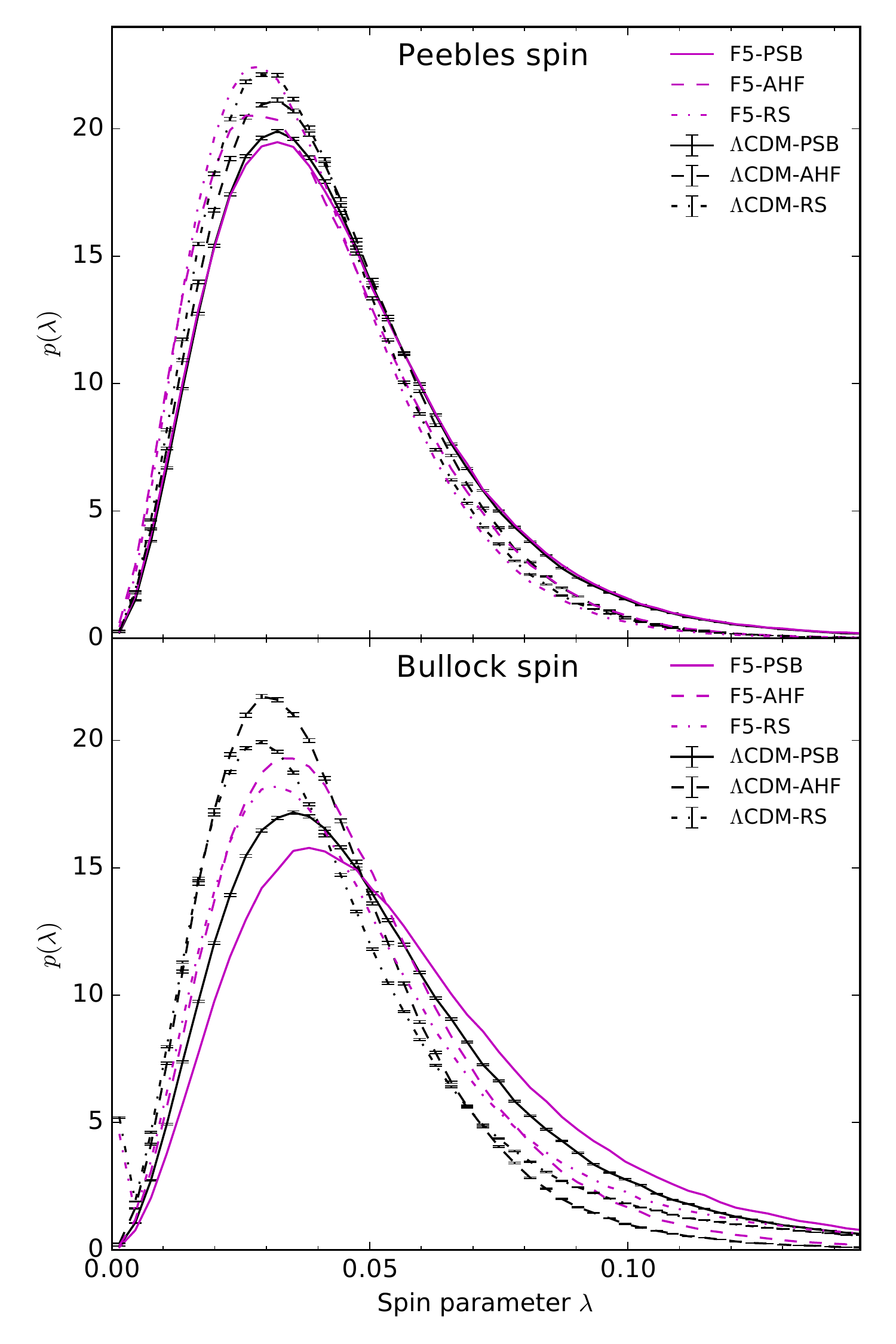}
  \caption{\label{fig:spin_PB}%
    PDF of the Peebles (top)  and Bullock (bottom) spin parameters for
f1024 GR (black) and F5 (magenta) at $z=0$, and for \code{PSB} (solid), \code{AHF} (dashed), and \code{Rockstar} (dash-dotted) lines. 
  }
\end{figure}

In  this  section,  we show how the definition of the spin parameter is affected by modified gravity. 
   \citet{1969ApJ...155..393P} defined the spin parameter as
\begin{equation}
  \label{eq:spin}
  \lambda_\text{P} = \frac{|\vect{J}|\sqrt{|E|}}{GM^{5/2}},
\end{equation}
where   $E =  W +  K$ is  the total  (kinetic plus  potential)
energy.\footnote{Note that in case of  MG, we calculated $E$ assuming
  GR.}

Fig.  \ref{fig:spin_PB} compares the  distributions of the Peebles and
Bullock's parameters in modified gravity.
In the Peebles definition, the energy is calculated assuming GR, which
underestimates the actual  energy of the halo, yielding  a lower value
of the spin parameter in MG.

We also show the effect of the halo-finder on the spin computation. 
For that purpose, we used  \code{AHF}  \citep{2004MNRAS.351..399G,
  2009ApJS..182..608K} to detect spherically overdense haloes
Moreover, the difference in the Peebles parameter between \code{PSB} and \code{AHF}/\code{Rockstar} can be understood by the following consideration.
\code{PSB} assumes a Plummer gravitational potential while calculating the potential energy, which is used for the unbinding step as well as the computation of the spin parameter.
\code{AHF} and \code{Rockstar}, however, assume a Newtonian potential.
Regarding the difference in the Bullock spin, \code{PSB} defines the  radius of a subhalo  as a function of the mass only, via equation~\eqref{eq:virial}, while \code{AHF} and \code{Rockstar} define it as the distance to the furthest particle.
Regardless of the differences, the trend is consistent between both definitions of the spin parameter: $f(R)$ gravity yields a larger Bullock spin, and has little effect on the Peebles spin. 


\bsp	
\label{lastpage}
\end{document}